\documentclass[a4paper,journal=jpcahf,manuscript=article,layout=twocolumn]{achemso}

\pdfoutput=1
\usepackage{stmaryrd,amssymb,rotating}
\usepackage{amsmath}
\usepackage[english]{babel}
\usepackage{amsfonts}
\usepackage{graphicx}
\usepackage{here}

\newcommand{\seesd}{See \ref{fig:sd} for a detailed explanation of the figure.}

\title{Visualizing Basins of Attraction \\
for Different Minimization Algorithms}
\date{\today}

\author{Daniel Asenjo} 
\author{Jacob D. Stevenson}
\email{js850@cam.ac.uk}
\phone{+44 (0)1223 336530}
\author{David J. Wales}
\author{Daan Frenkel}

\affiliation{Department of Chemistry, University of Cambridge,
  Lensfield Road, Cambridge, CB2 1WE, United Kingdom}

\begin{document}

\begin{abstract}
  We report a study of the basins of attraction for potential energy minima
  defined by different minimisation algorithms for an atomic system. We find
  that whereas some minimisation algorithms produce compact basins, others
  produce basins with complex boundaries or basins consisting of disconnected
  parts. Such basins deviate from the `correct' basin of attraction defined
  by steepest-descent pathways, and the differences can be controlled to some
  extent by adjustment of the maximum step size. 
  The choice of the most convenient minimisation
  algorithm depends on the problem in hand. We show that while L-BFGS is the
  fastest minimiser, the FIRE algorithm is also quite fast, and can lead to
  less fragmented basins of attraction.
\end{abstract}


\section{Introduction}
\label{sec:intro}

Optimisation problems are ubiquitous in the physical sciences and
beyond. In the simplest case optimisation refers to the search for the
minimum or maximum values of an objective function. Global
optimisation involves searching for the highest maximum or lowest
minimum in a certain domain. In contrast, local optimisation
procedures identify the first minimum or maximum that is found by a
given algorithm when starting from an arbitrary point in parameter
space.

In the study of energy landscapes, the properties of stationary points
and the connections between them are of central
importance.\cite{Wales2003} These stationary points represent key
features of the landscape. In chemical reactions, saddle points are
geometric transition states \cite{murrelll68} along the reaction
coordinate. Glassy systems are trapped in metastable states that
correspond to relatively small numbers of connected local minima
\cite{kirkpatricktw89,DoliwaH03,deSouzaW08} and, similarly, jammed
states can also be viewed as local potential energy minima.  In
protein folding the native state corresponds to the global free energy
minimum and the free energy landscape often involves funnelling
characteristics.\cite{Wolynes1995} The study of these minima and the
pathways connecting them can be carried out using geometry
optimisation techniques,\cite{Wales2003,WalesB06} and local
minimisation is the focus of the current contribution.

When faced with the task of numerically optimising a smooth function
there are many algorithms from which to choose. Which algorithm is
best suited for the purpose depends on factors such as speed, memory
usage and ease of implementation.  All algorithms follow a general
procedure starting with the user supplying an initial point, which can
be an informed guess or an arbitrary point in parameter space. The
algorithm generates a sequence of iterates that terminates when a
solution is found within a predefined accuracy, such as when the
gradient is near zero, or when the value of the function stops
changing.  Recent work has shown how convergence criteria can be
chosen according to a certification procedure.\cite{MehtaHW13}
Different algorithms have different ways of proceeding from one
iteration to the next. The formulations we consider involve the value
of the function that is being optimised, its derivatives, and the
results of previous iterations.

In general, two different algorithms can converge to different minima
starting from the same initial conditions.  We are interested in
identifying the configuration space that leads to a particular minimum
as a function of the optimisation algorithm.  This connection is
important for applications such as calculation of thermodynamic
properties using the superposition approach, where the global
partition function is written as a sum over contributions of local
minima.\cite{wales93f,doyew95c,StrodelW08b} In this context, the
steepest-descent algorithm occupies a unique position, since it
defines basins of attraction\cite{mezey81b} for local minima that
cannot interpenetrate. This result follows because steepest-descent
paths are defined using a linear first-order differential equation,
for which the uniqueness theorem applies.\cite{pechukas76} However,
steepest-descent minimisation is very inefficient compared to more
sophisticated algorithms, which are normally preferable.  The latter
methods generally employ non-linear equations to determine the steps,
and the corresponding basins of attraction can exhibit complex
boundaries.\cite{Wales1992,Wales1993} In other words, when defined by
steepest-descent, the basin of attraction has a deterministic
boundary. In contrast, the basins for other algorithms can exhibit
re-entrant, interpenetrating boundaries, as noted in previous
work.\cite{Wales1992,Wales1993} For applications such as basin-hopping
global optimisation\cite{walesd97a,lis87} this structure is probably
unimportant. However, the simplicity of the boundaries associated with
steepest-descent paths is relevant if we are interested in
partitioning the configuration space, for example, when measuring the
size of basins of attraction. \cite{xu13} In the present work, we
regard the basins of attraction defined by steepest descent as a
reference to which other methods will be compared.  The purpose of
this paper is to make these comparisons rigorously and to provide
criteria for choosing the most appropriate and convenient minimisation
algorithm for a given problem.

\section{Methods}
\label{sec:methods}

The minimisation algorithms considered here are steepest-descent,
L-BFGS, FIRE, conjugate gradient, and BFGS (for a detailed description
of different minimisation techniques see
reference\cite{Nocedal1999}). The L-BFGS algorithm is tested using two
different methods to determine the length of the steps. In the first
approach, a line search routine is used to choose the step size.  In
the second approach, the step size guess of the L-BFGS algorithm is
accepted subject to a maximum step size and the condition that the
energy does not rise.  This is the default procedure in the global
optimisation program {\tt GMIN}\cite{gmin} and the {\tt OPTIM} program
for locating transition states and analysing pathways.\cite{optim} We
have only compared gradient-based minimisers in the present work,
because they represent the most efficient class of algorithms.

Steepest-descent, sometimes referred to as gradient descent, uses the
gradient as the search direction (this is the steepest direction). The
step size can be chosen using a line search routine. In this paper, a
fixed step size ($\Delta=0.005$ in reduced units) is used for all of
the steepest-descent calculations. It is worth noting that the
definition of basins of attraction in the Introduction section applies
to steepest-descent minimisation in the limit of infinitesimal step
size.

BFGS, named after its creators Broyden,\cite{Broyden1970}
Fletcher,\cite{Fletcher1970} Goldfarb,\cite{Goldfarb1970} and
Shanno,\cite{Shanno1970} is a quasi-Newtonian optimisation method,
which uses an approximate Hessian to determine the search
direction. The approximate Hessian is built up iteratively from the
history of steps and gradient evaluations.  The implementation used in
this paper is from SciPy\cite{scipy} and uses a line search to
determine a step size.  The line search used is the Minpack2 method
DCSRCH,\cite{more.1994} which attempts to find a step size that
satisfies the Wolfe conditions.  The maximum step size is fixed to be
$50$ times the initial guess returned by the BFGS algorithm.

L-BFGS is a limited memory version of the BFGS algorithm described
above and was designed for large-scale problems, where storing the
Hessian would be impractical. Rather than saving the full approximate
Hessian in memory it only stores a history of $M$ previous values of
the function and its gradient with which it computes an approximation
to the inverse diagonal components of the Hessian.\cite{Nocedal1989}
For a system with $N$ variables, ${\cal O}(N^2)$ memory and operations
are needed when using BFGS, while L-BFGS scales as ${\cal O}(MN)$,
which is significantly smaller if $M\ll N$, and is linear in $N$. Two
versions of L-BFGS were used in this paper. The first is from the
SciPy\cite{scipy} optimisation library
"L-BFGS-B".\cite{Zhu1997,Byrd1995,Morales2011} This routine uses the
same DCSRCH line search as the BFGS implementation, but with slightly
different input parameters. For example, the maximum step size is
adaptively updated.  The second L-BFGS implementation is included in
the {\tt GMIN}\cite{gmin} and {\tt OPTIM}\cite{optim} software
packages and adapted from Liu and Nocedal.\cite{Nocedal1989} In this
version there is no line search. The step size returned by the L-BFGS
algorithm is accepted subject to a maximum step size constraint and
the condition that the energy does not rise.  In both of these
versions, the diagonal components of the inverse Hessian are initially
set to unity.

The fast inertial relaxation engine, known as FIRE, is a minimisation algorithm
based on ideas from molecular dynamics, with an extra velocity term and
adaptive time step.\cite{Bitzek2006}  Stated simply, the system state slides down
the potential energy surface gathering ``momentum'' until the direction of the
gradient changes,  at which point it stops, resets, and resumes sliding.

The conjugate gradient method uses information about previous values of the
gradient to determine a conjugate search direction.\cite{Stiefel1952} It only
stores the previous search direction.  The implementation considered here is from
SciPy,\cite{scipy} and the step size is determined using same line search as
the BFGS routine.

In order to test the accuracy of the minimisers, we use a
three-particle system in which the inter-particle interactions are
given by a Lennard-Jones potential plus a three-body Axilrod--Teller
term:\cite{LJ1924,Axilrod1943}
\begin{eqnarray}
\label{eq:pot}
V&=&4\varepsilon \sum_{i<j}\left[\left(\frac{\sigma}{r_{ij}}\right)^{12} 
- \left(\frac{\sigma}{r_{ij}}\right)^6 \right] +\\ 
& &\qquad Z\sum_{i<j<k}\left[
\frac{1+3\cos\theta_1\cos\theta_2\cos\theta_3}{(r_{ij}r_{ik}r_{jk})^3} \right],
   \nonumber
\end{eqnarray}
Here $\theta_1$, $\theta_2$ and $\theta_3$ are the internal angles of
the triangle formed by particles $i$, $j$, $k$; $r_{ij}$ is the
distance between particles $i$ and $j$; and $Z$ is the strength of the
three-body term. We chose this three-particle system because, for
$Z>0$, it has four local minima.  It was important for us to choose a
small system with only a few degrees of freedom in order to visualise
the basins of attraction in two dimensions.  In one of the minima, the
atoms are arranged in an equilateral triangle, with point group
$D_{3h}$.  The other three linear minima have $D_{\infty h}$ symmetry
and are related by permutations of the atoms.  We use reduced units
with one parameter, as in previous
work:\cite{wales90a,doyew92,Wales1992,Wales1993}
\begin{equation}
\label{eq:z}
Z^*=\frac{Z\sigma^9}{\varepsilon}.
\end{equation}

\begin{figure}
  \centering
  \includegraphics[width=0.4\textwidth]{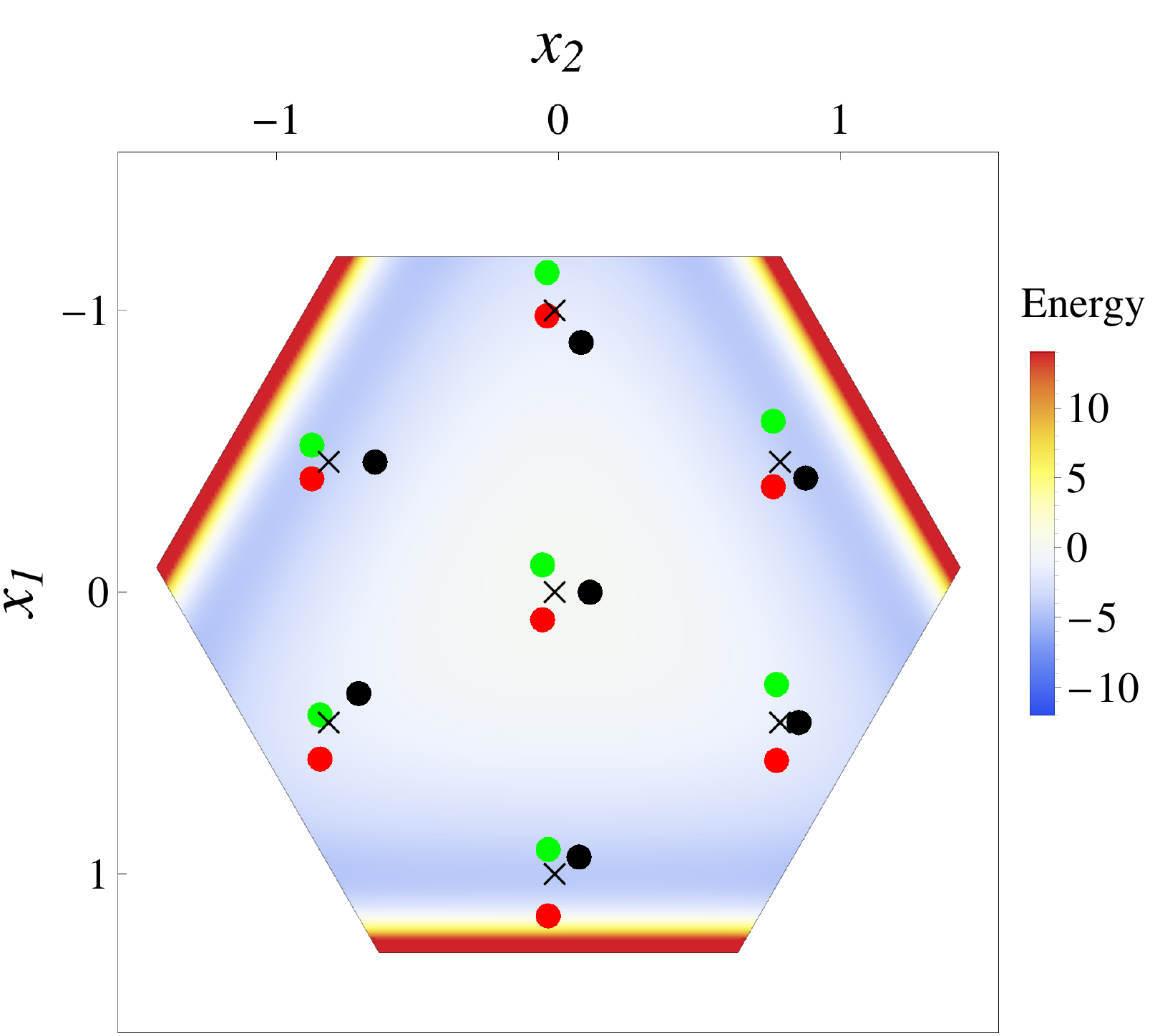} 
  \caption{Projected plane in its own coordinate system.  Points on
    this plane were used as the initial configurations for the
    minimisations. Examples of starting configurations are shown for
    several points marked by ``$\times$''.  The plot is coloured by the
    energy of the starting configuration.}
  \label{fig:projection}
\end{figure}

Without loss of generality, we define the axes such that the three
particles are in the $xy$ plane with one particle at the origin,
another along the $x$ axis, and the third in the upper half plane. Now
only the three internal coordinates $r_{ij}$ are needed to describe
the system.  A projection onto the page was chosen to visualise the
basins of attraction in such a way that the basins of the four minima
are present in the plane.\cite{Wales1992,Wales1993} In internal
coordinates, the projection plane is chosen to be perpendicular to the
vector $\vec n = (1,1,1)$ at a distance $\sqrt{3}\alpha$ from the
origin.  Points in the plane have the property
$r_{12}+r_{23}+r_{13}=3\alpha$.  We can define an arbitrary vector
$\vec v = (0,0,1)$ so that the plane is spanned by the unit vectors
\begin{equation}
\hat x_2 = \frac{\vec n \times \vec v}{|\vec n \times \vec v|}, \qquad
\hat x_1 = \frac{\vec n \times \vec x_2}{|\vec n \times \vec x_2|}.
\end{equation}
The projection of an arbitrary vector $\vec a =
(r_{12},r_{23},r_{13})$ onto the plane is $\vec a_p = (\vec a \cdot
\hat x_1, \vec a \cdot \hat x_2)$.  The equilateral triangle minimum
is at the origin in terms of the projected coordinates $x_1$ and
$x_2$, as shown in \ref{fig:projection}. For more details regarding
the projection see references \cite{Wales1992,Wales1993}.

The following results were produced using the projection described
above with $\alpha=\sqrt{3}R_e$, where $R_e=2^{1/6}\sigma$ is the
Lennard-Jones equilibrium separation, and $Z^*=2$.  For this choice
there is a linear minimum with energy $-2.219\,\varepsilon$ in
addition to the equilateral triangle with energy
$-2.185\,\varepsilon$.  There are three distinct permutational isomers
of the linear minimum, since any of the three atoms can reside in the
central position.  A $700\times 700$ grid of initial points was taken
with $x_1$ and $x_2$ between $-\alpha$ and $\alpha$. All of the
minimisations were terminated when the root mean square (RMS) gradient
was smaller than $10^{-3}$ reduced units. Under some conditions, such
geometry optimisations could appear to converge to a saddle
point,\cite{uppenbrinkw92b} so the geometries were also checked, as
well as the RMS gradient.  As in previous
work,\cite{Wales1992,Wales1993} each pixel in the resulting plots
corresponds to a different initial configuration and is coloured
according to the minimum that is found after optimisation.

The efficiency of each algorithm was also tested.  Here we were not
constrained to small systems by the need for visualisation, so we
chose a more interesting system size, namely 38 Lennard-Jones atoms.
We measured the average number of function calls needed to get to the
nearest local minimum from 1,000 random starting configurations. The
number of function calls is a fairer test than wall clock time because
for most real word calculations computing the energy and gradient will
be the time bottleneck and it avoids measuring differences in
implementation efficiency.  The results are reported in \ref{tab:err}.

\section{Results and Discussion}
\label{sec:results}

\begin{figure}
  \centering \includegraphics[width=0.4\textwidth]{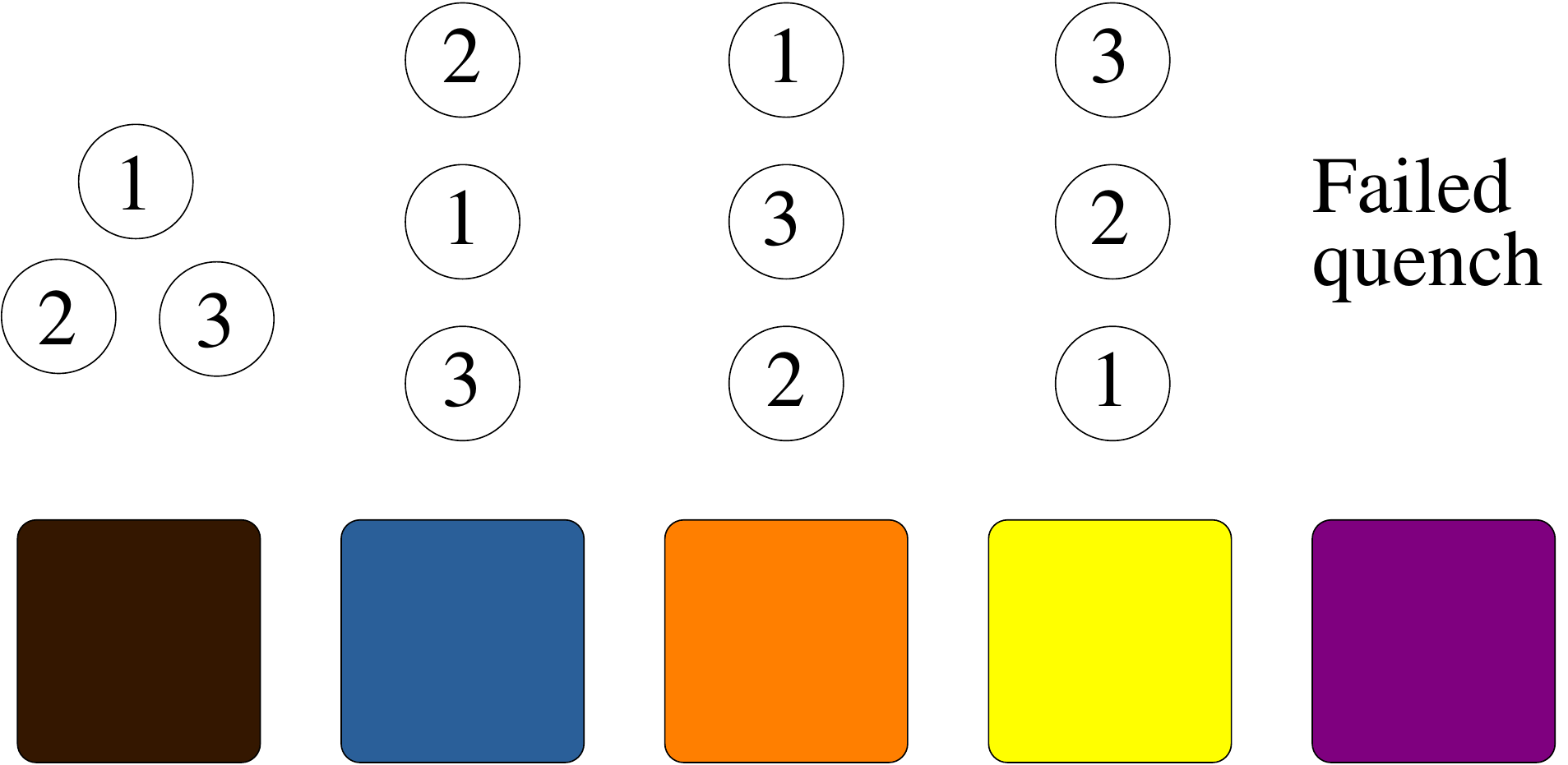}\\
  \vspace{.5cm} Steepest-descent, $\Delta=0.005$ \\ 
  \includegraphics[width=0.4\textwidth]{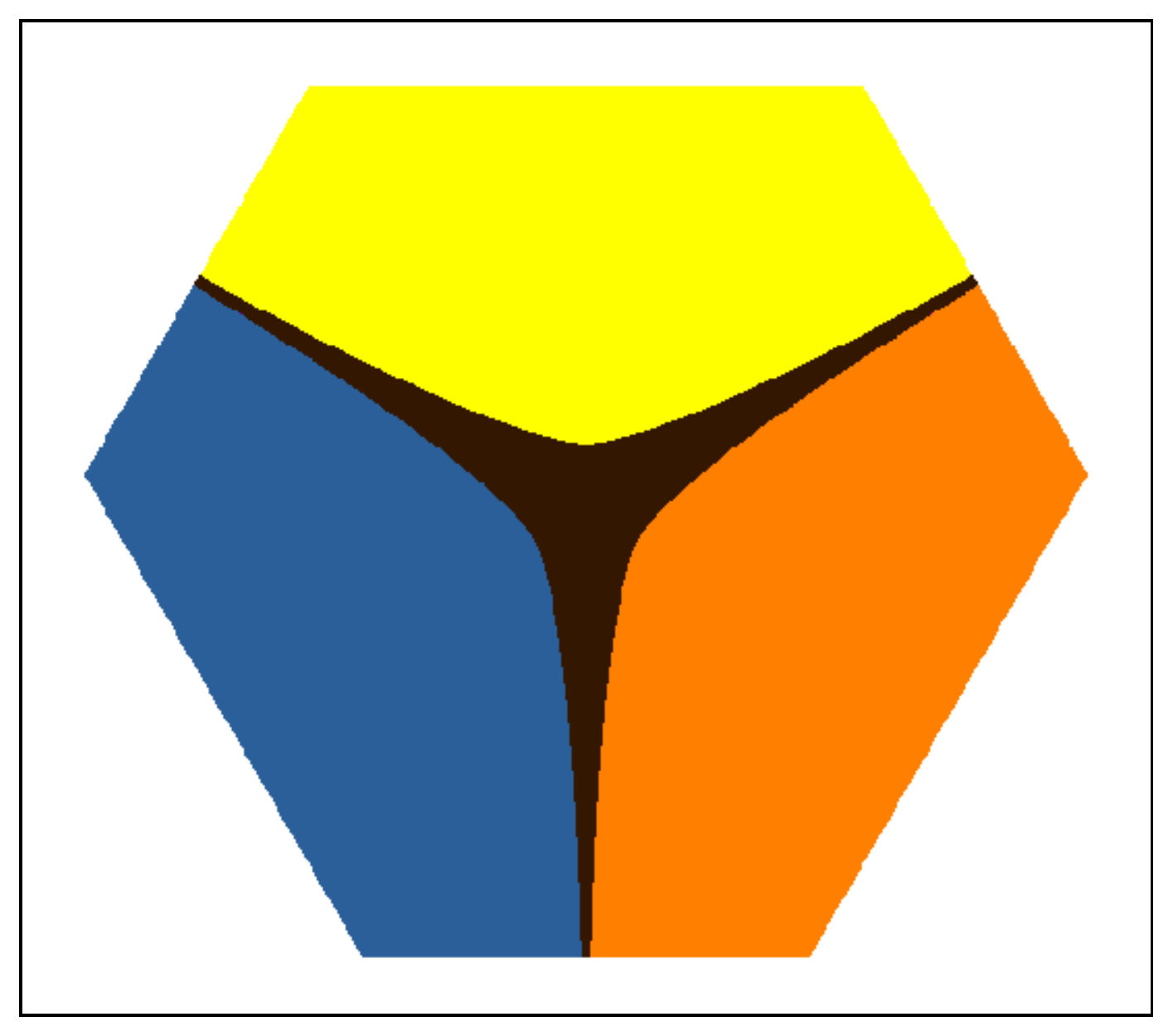} 
  \caption{ The lower panel shows the same $(x_1, x_2)$ plane as
    \ref{fig:projection}, which defines the starting configuration of
    the three particles that interact via the Lennard-Jones potential plus
    a three-body Axilrod-Teller term, as described in the Methods
    Section.  The plane is coloured according to the final
    configuration after a steepest-descent minimisation.  The colour
    coding is displayed in the upper panel.  Black corresponds to the
    atoms in the $D_{3h}$ triangle configuration, while blue, orange, and yellow
    correspond to the three possible linear configurations.  A failed
    quench means that the final coordinates are not close enough
    (according to a certain tolerance) to the equilateral triangle or
    linear configurations.  $\Delta$ is the maximum step size.  This
    figure, corresponding to the steepest-descent results, will
    serve as our reference for comparing the other minimisers.}
  \label{fig:sd}
\end{figure}

\ref{fig:sd} shows the colour scheme used to identify the results of
local minimisation in the subsequent figures. Forbidden geometry
refers to the points in the plane that correspond to initial
geometries that do not satisfy the triangle inequality or have
excessively high energy. Failed quench means that the quenched
coordinates are not close enough (according to a certain tolerance) to
the equilateral triangle or linear configurations, that is, the
algorithm failed to reach a minimum.\cite{uppenbrinkw92b}

The figures that follow show the basins of attraction of the four
minima described above determined using the different minimisation
techniques and parameters, as described in the Methods section.  As
expected, steepest-descent is the slowest (see \ref{tab:difpix}), most
robust minimiser, and it produces well-defined basin boundaries
(\ref{fig:sd}).  This result holds as long as the step size is kept
relatively small. Smaller step sizes are always more robust when using
steepest-descent.  The usual definition for the basin of attraction in
the context of energy landscapes is the set of points in configuration
space that converge to a certain minimum for a steepest-descent
quench.\cite{mezey81b,Wales2003} Hence this approach produces a useful
reference against which to compare the other algorithms.

\begin{figure*}
  \centering 
  L-BFGS, $M=1$, $\Delta=0.05$ \hfill L-BFGS, $M=1$,
  $\Delta=0.1$ \hfill L-BFGS, $M=1$, $\Delta=0.2$
  \includegraphics[width=0.3\textwidth]{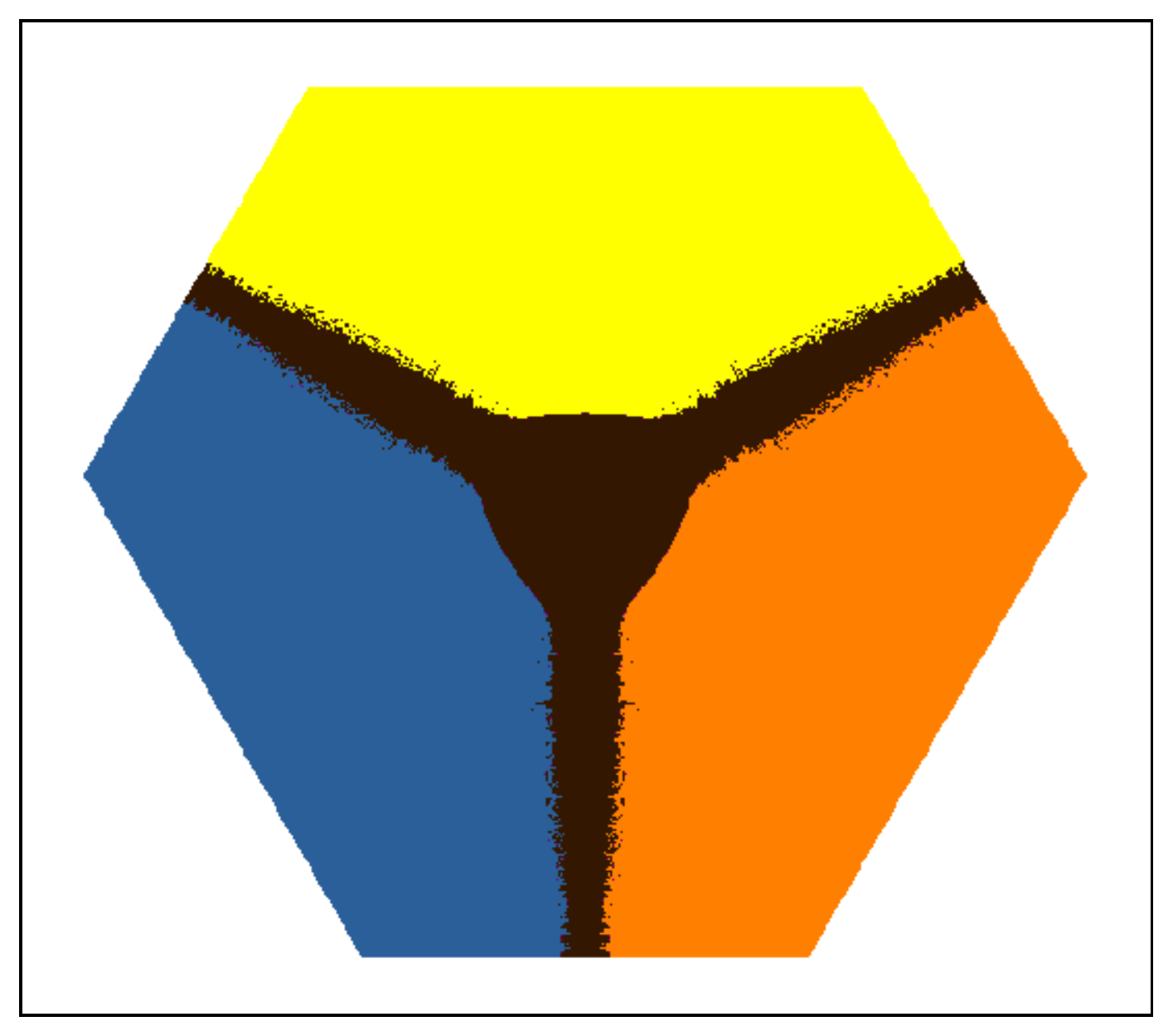}\hfill
  \includegraphics[width=0.3\textwidth]{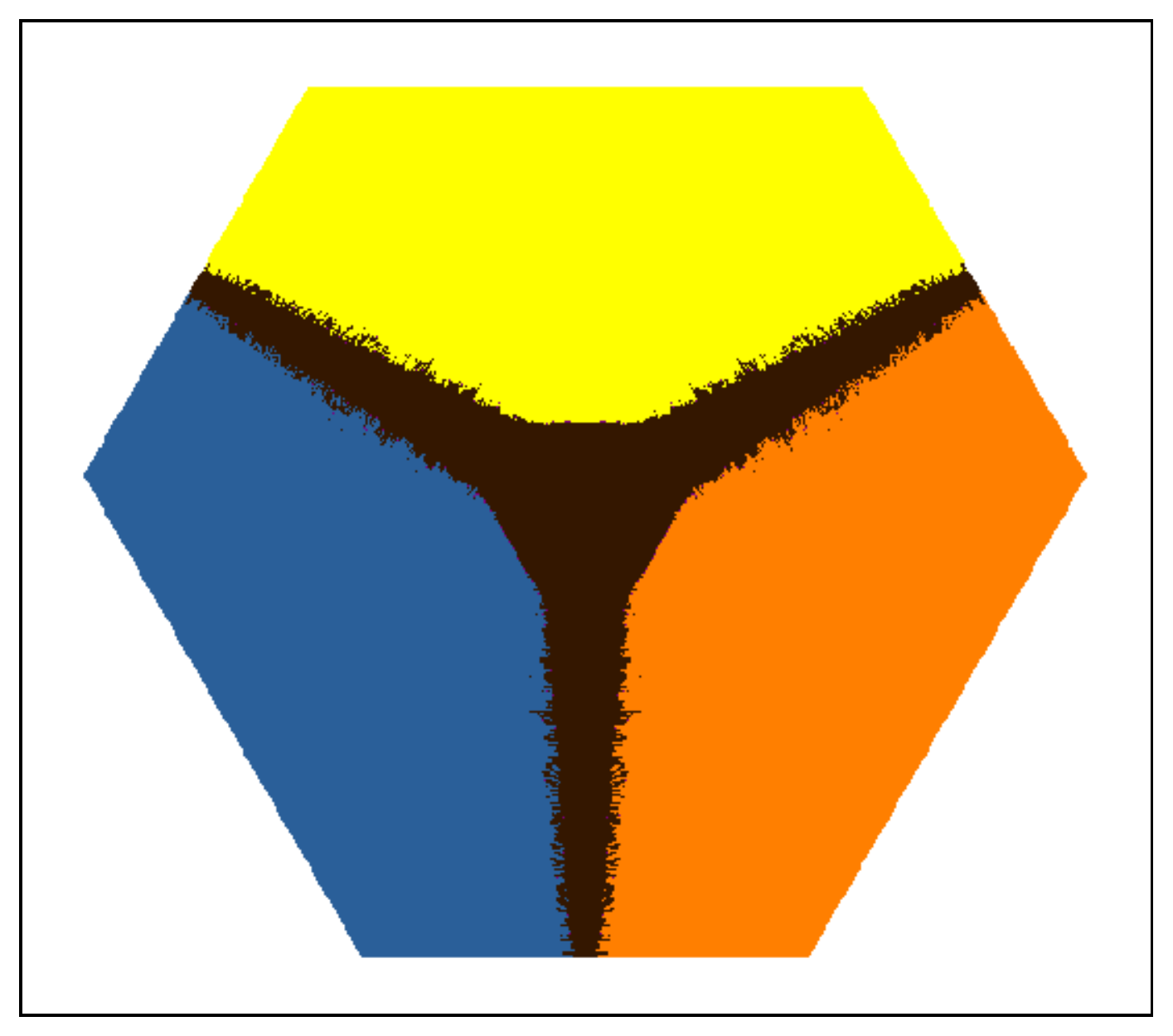}\hfill
  \includegraphics[width=0.3\textwidth]{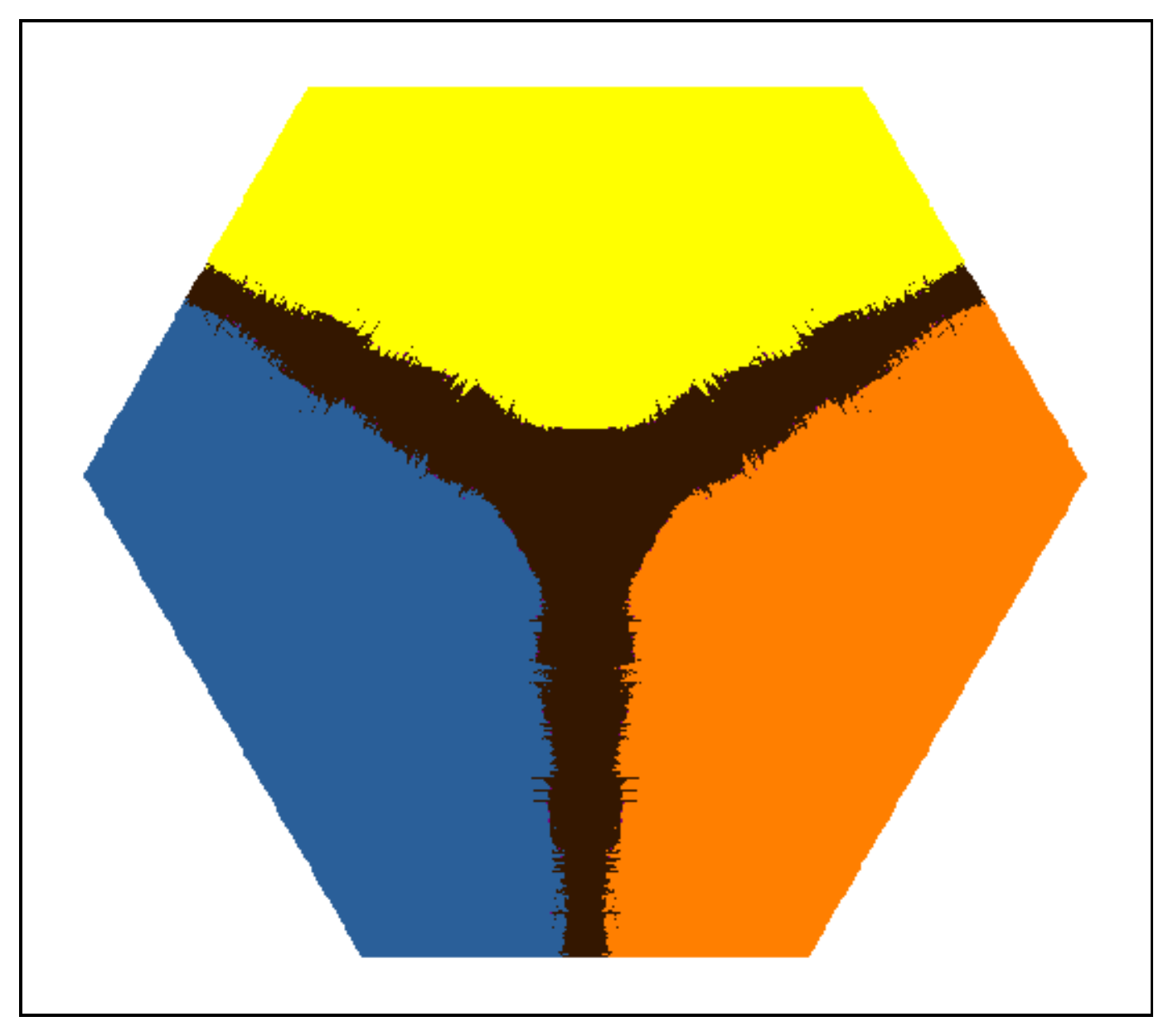} \\
  L-BFGS, $M=4$, $\Delta=0.05$ \hfill L-BFGS, $M=4$,
  $\Delta=0.1$ \hfill L-BFGS, $M=4$, $\Delta=0.2$
  \includegraphics[width=0.3\textwidth]{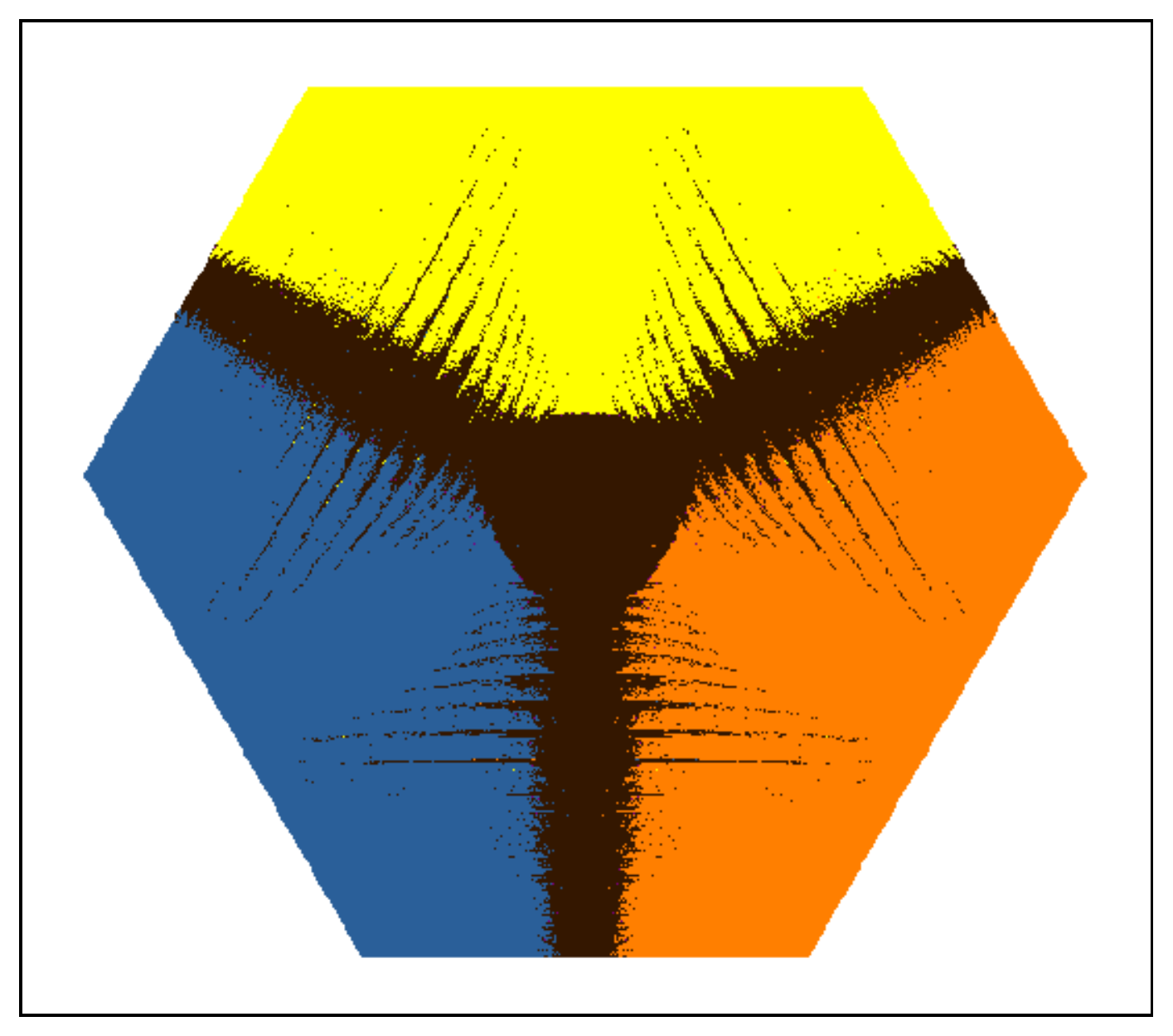}\hfill
  \includegraphics[width=0.3\textwidth]{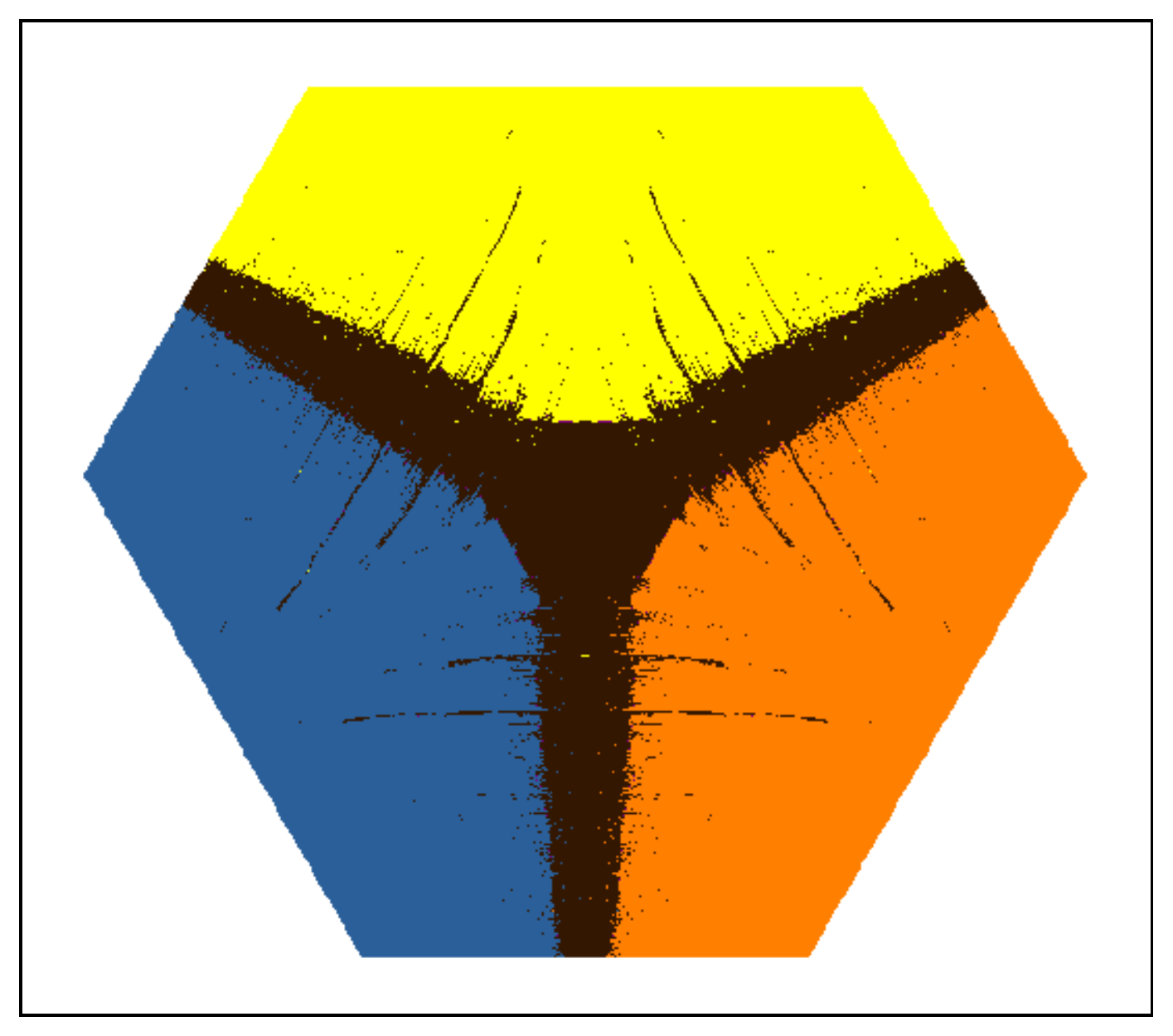}\hfill
  \includegraphics[width=0.3\textwidth]{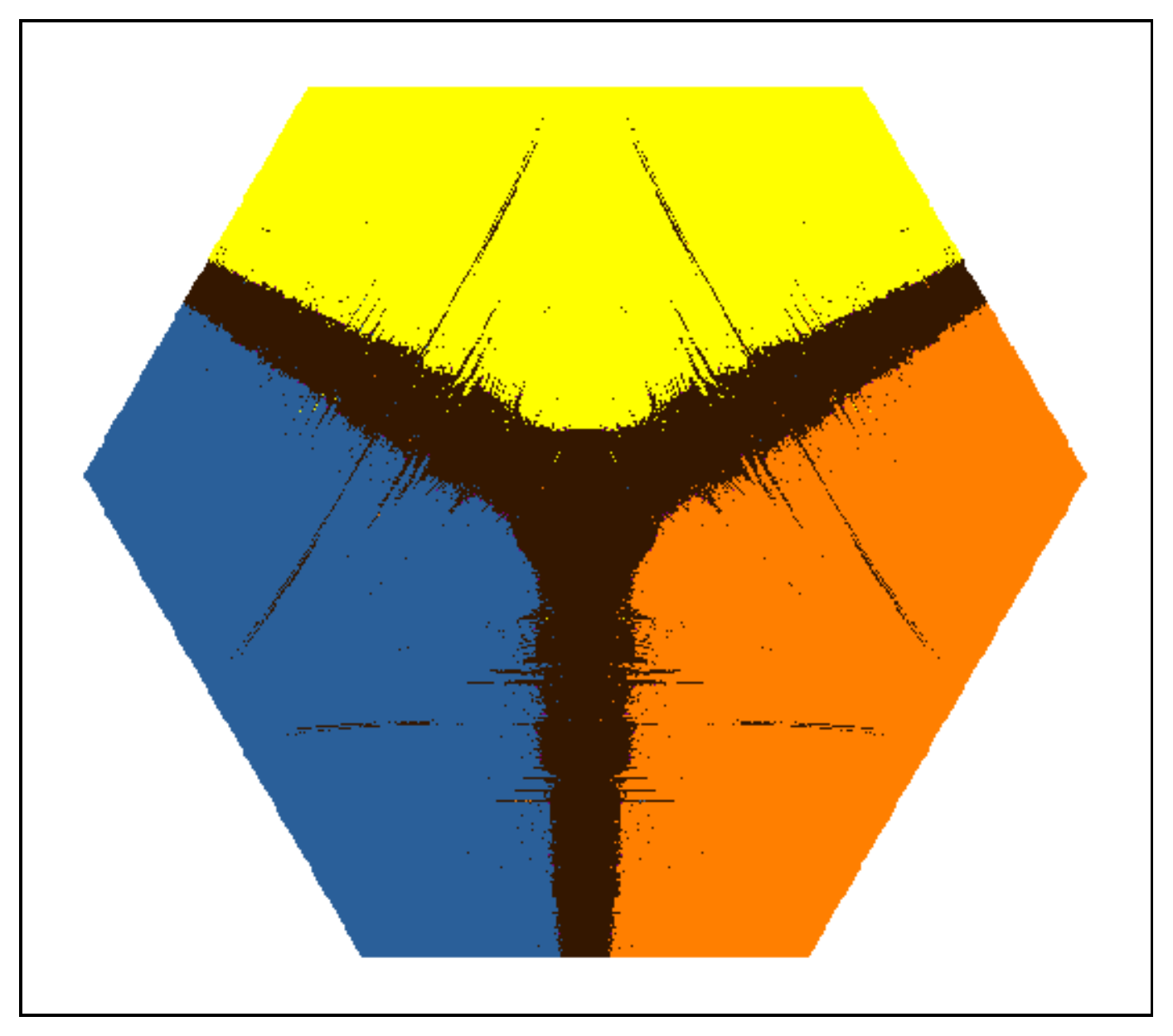} \\
  L-BFGS, $M=10$, $\Delta=0.05$ \hfill L-BFGS, $M=10$,
  $\Delta=0.1$ \hfill L-BFGS, $M=10$, $\Delta=0.2$
  \includegraphics[width=0.3\textwidth]{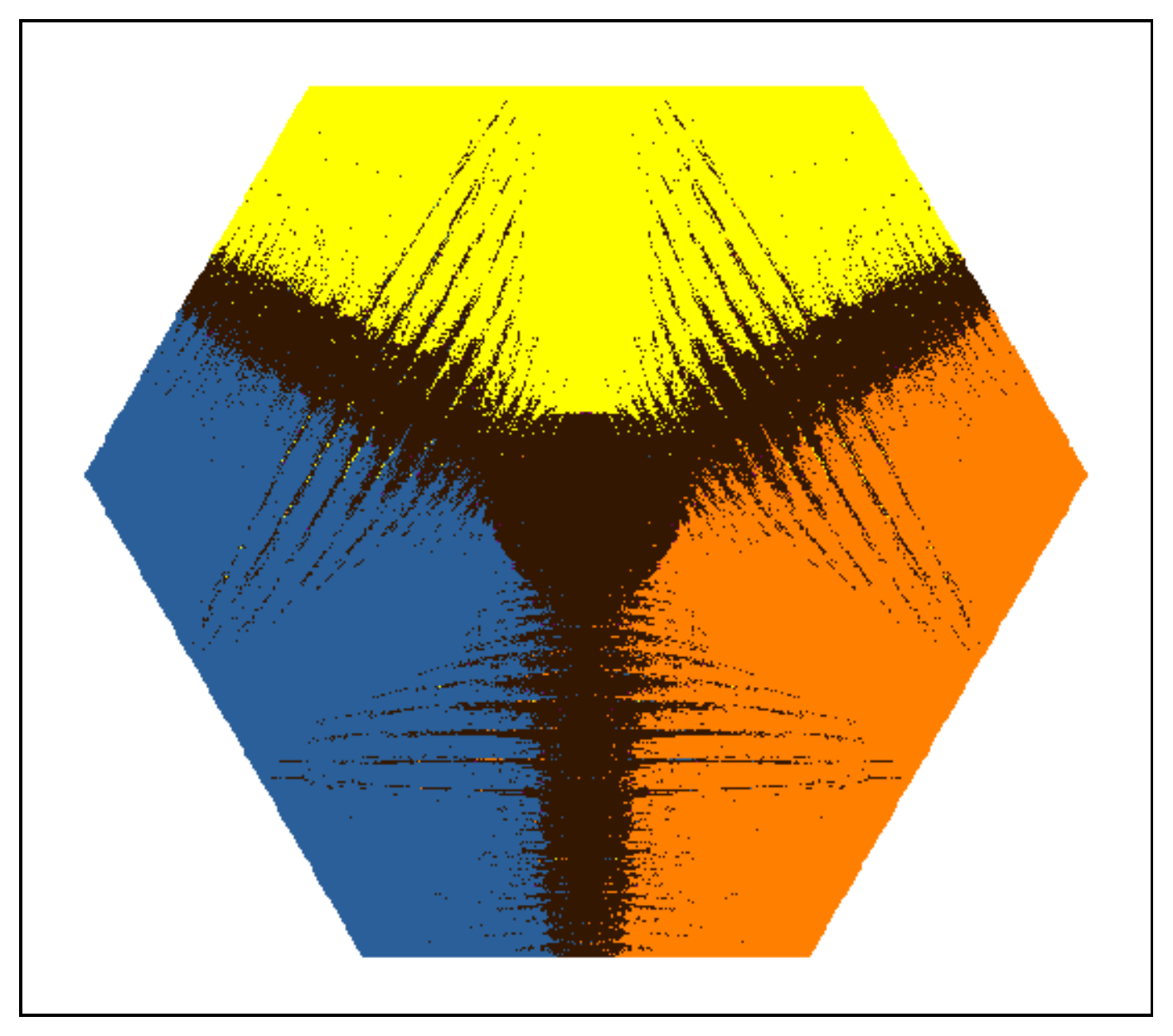}\hfill
  \includegraphics[width=0.3\textwidth]{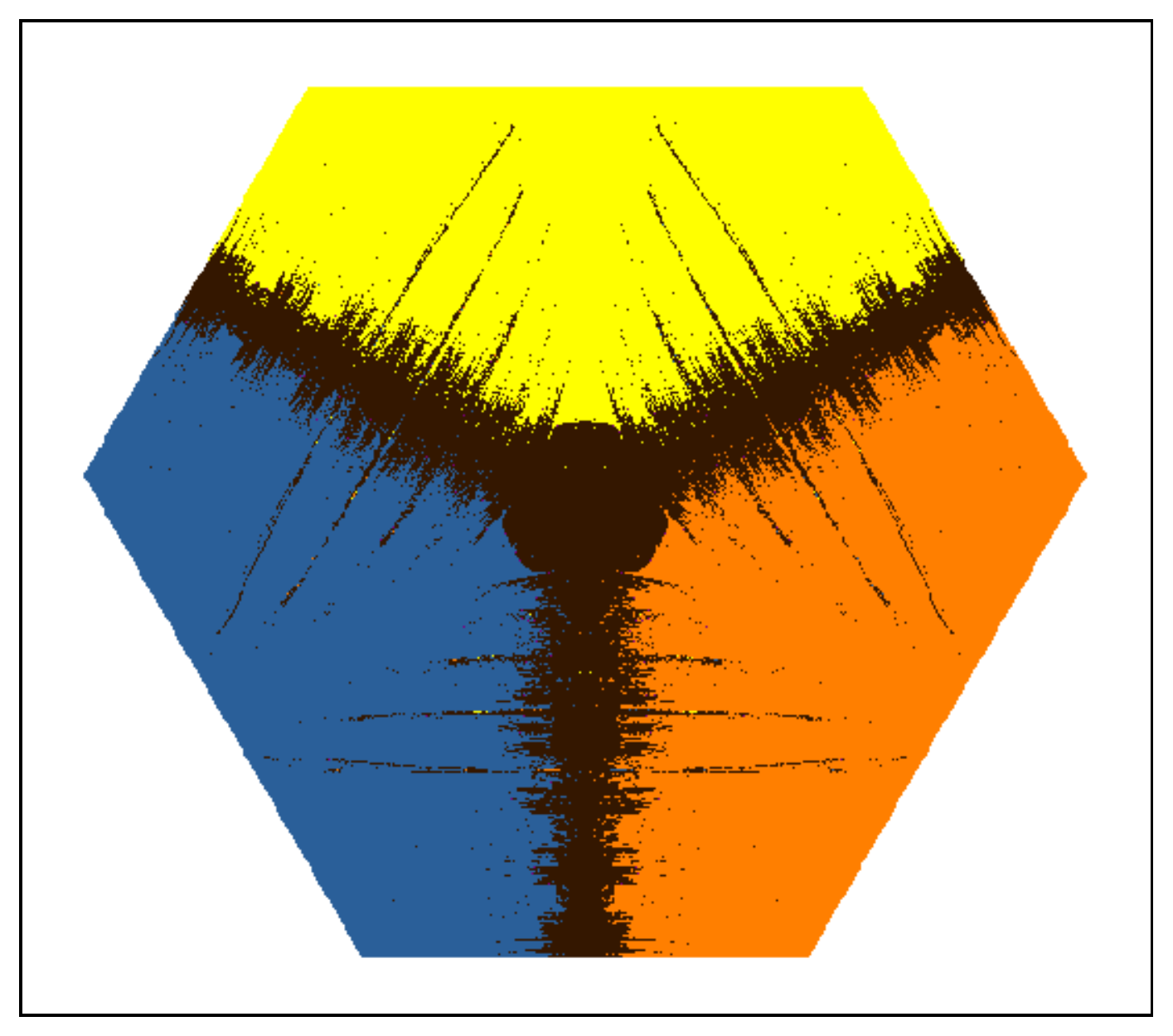}\hfill
  \includegraphics[width=0.3\textwidth]{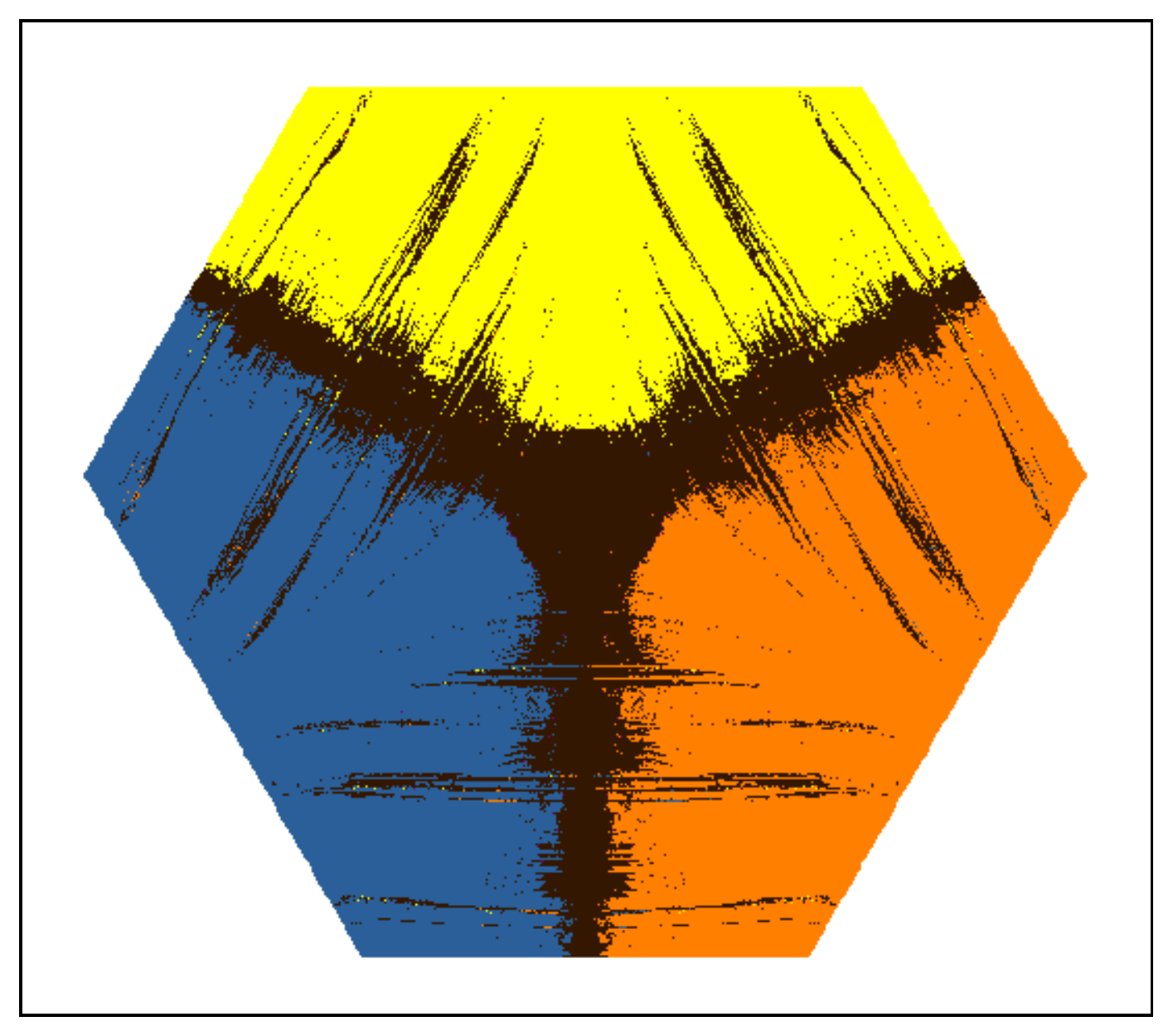} \\
  \caption{Results for the L-BFGS algorithm for different values of $M$ (the
    number of previous steps used to construct the next step) and the maximum
    step size $\Delta$. \seesd}
  \label{fig:lbfgs}
\end{figure*}

\begin{figure*}
  \centering
  L-BFGS, with line search (SciPy) \hspace{.1\textwidth} BFGS, with line search (SciPy) \\
  \includegraphics[width=0.4\textwidth]{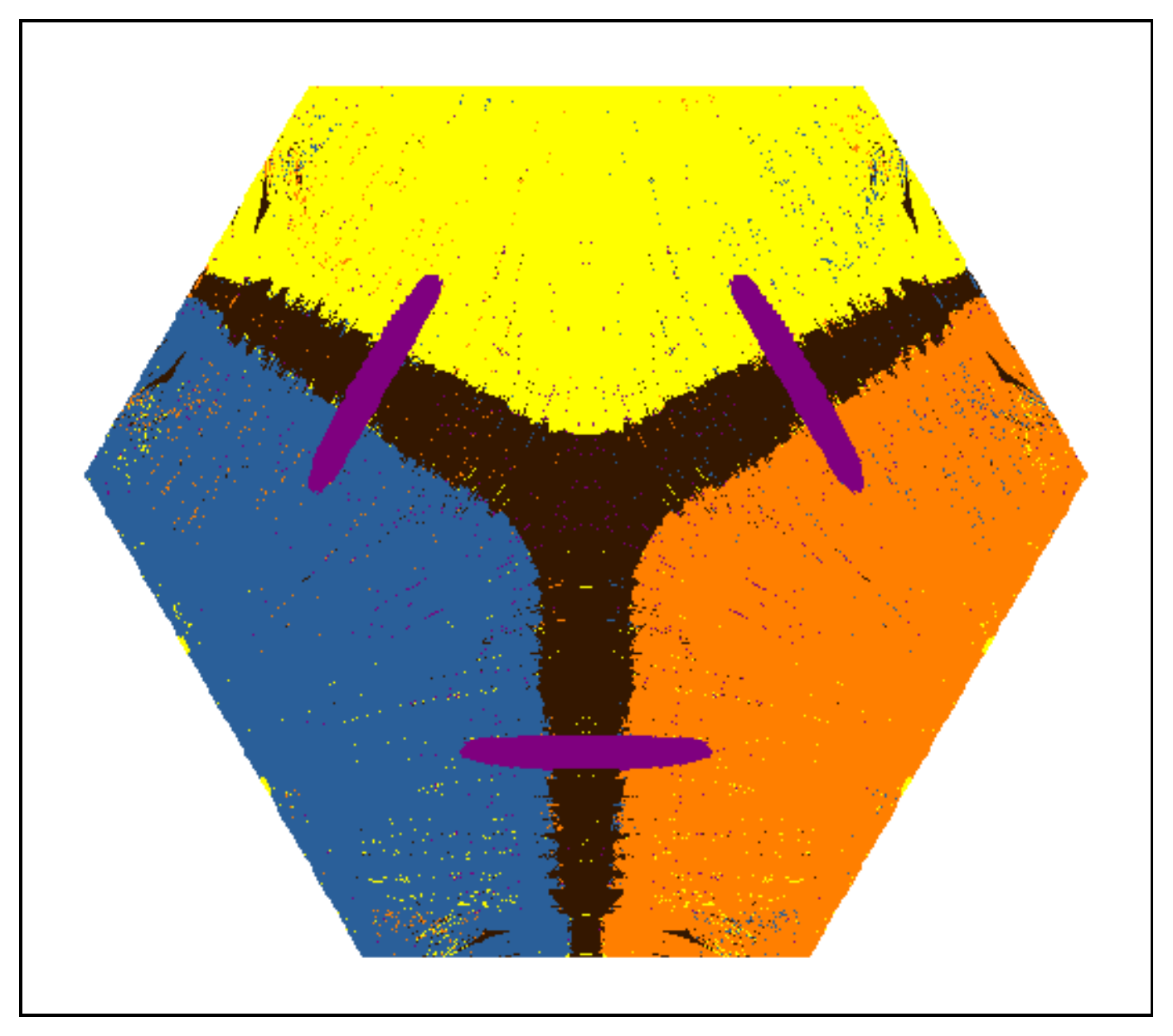}
  \includegraphics[width=0.4\textwidth]{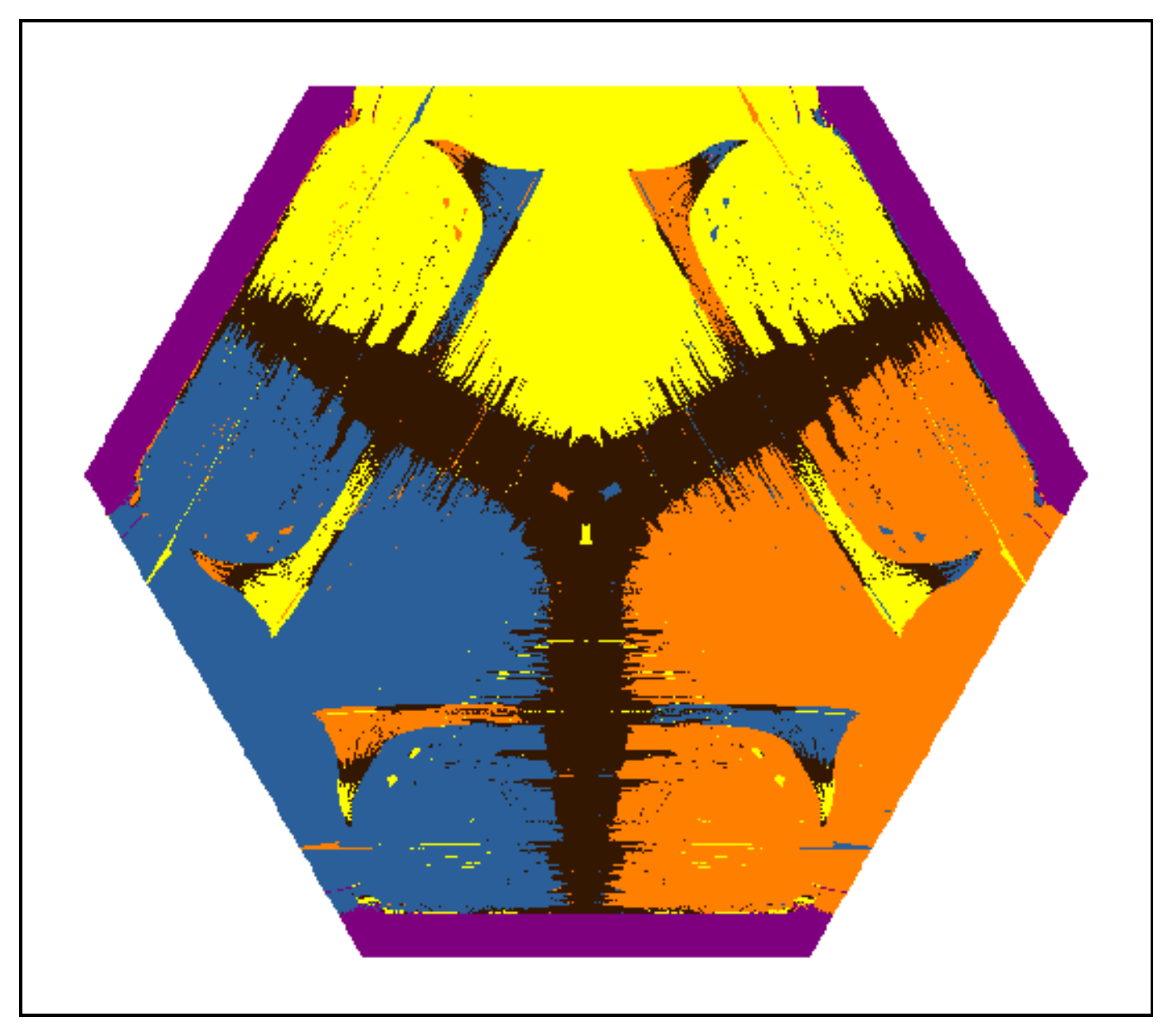}
  \caption{Results for L-BFGS and BFGS with line search (SciPy implementations).
    \seesd
  }
  \label{fig:bfgs}
\end{figure*}

L-BFGS is the fastest algorithm tested here (see \ref{tab:err}),
although the basin boundaries are not always well defined (see
\ref{fig:lbfgs} and \ref{fig:bfgs}). In the case of L-BFGS without
line search we can see that reducing the step size does not
necessarily improve the definition of the basin boundaries.  In this
case, the resolution of the basin boundaries improves with increasing
maximum step size until it reaches an optimum length, beyond which the
resolution decreases. This effect is clearly visible in
\ref{tab:difpix} and \ref{fig:lbfgs}. Removing the line search does
not improve the resolution of the boundaries, but it does reduce the
number of failed quenches. We tested the effect of changing the
parameter $M$, the number of previous values of the function and
gradient used to build the approximate Hessian. Increasing $M$ between
1 and 10 makes the resolution of the basins worse (see
\ref{fig:lbfgs}) but produces faster convergence (see
\ref{tab:difpix}). This result arises due to the fact that increasing
$M$ increases the degree of non-linearity of the algorithm.

\begin{figure*}
  \centering FIRE, $\Delta=0.1$ \hspace{.25\textwidth} FIRE,
      $\Delta=0.25$\\
      \includegraphics[width=0.4\textwidth]{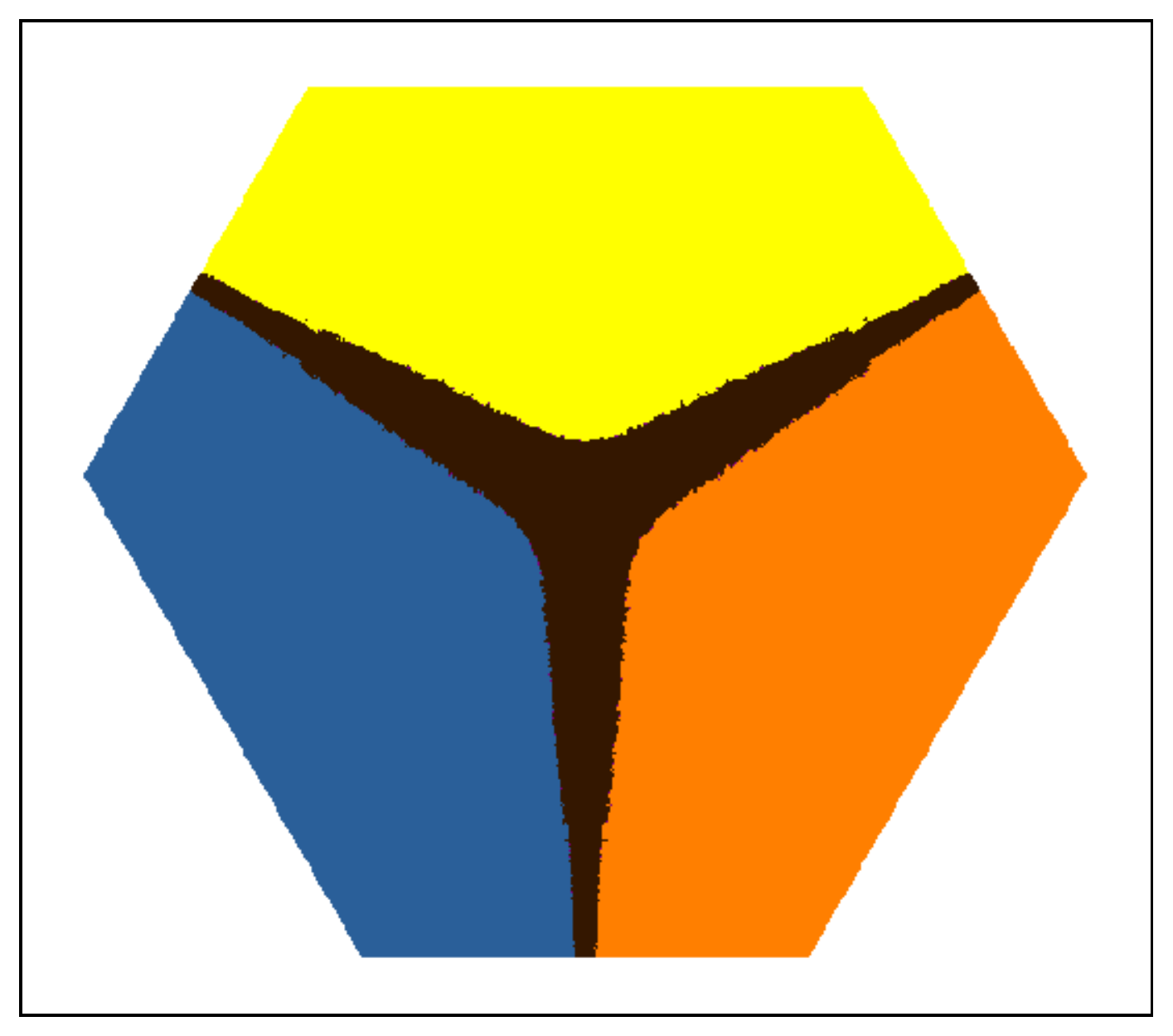}
      \includegraphics[width=0.4\textwidth]{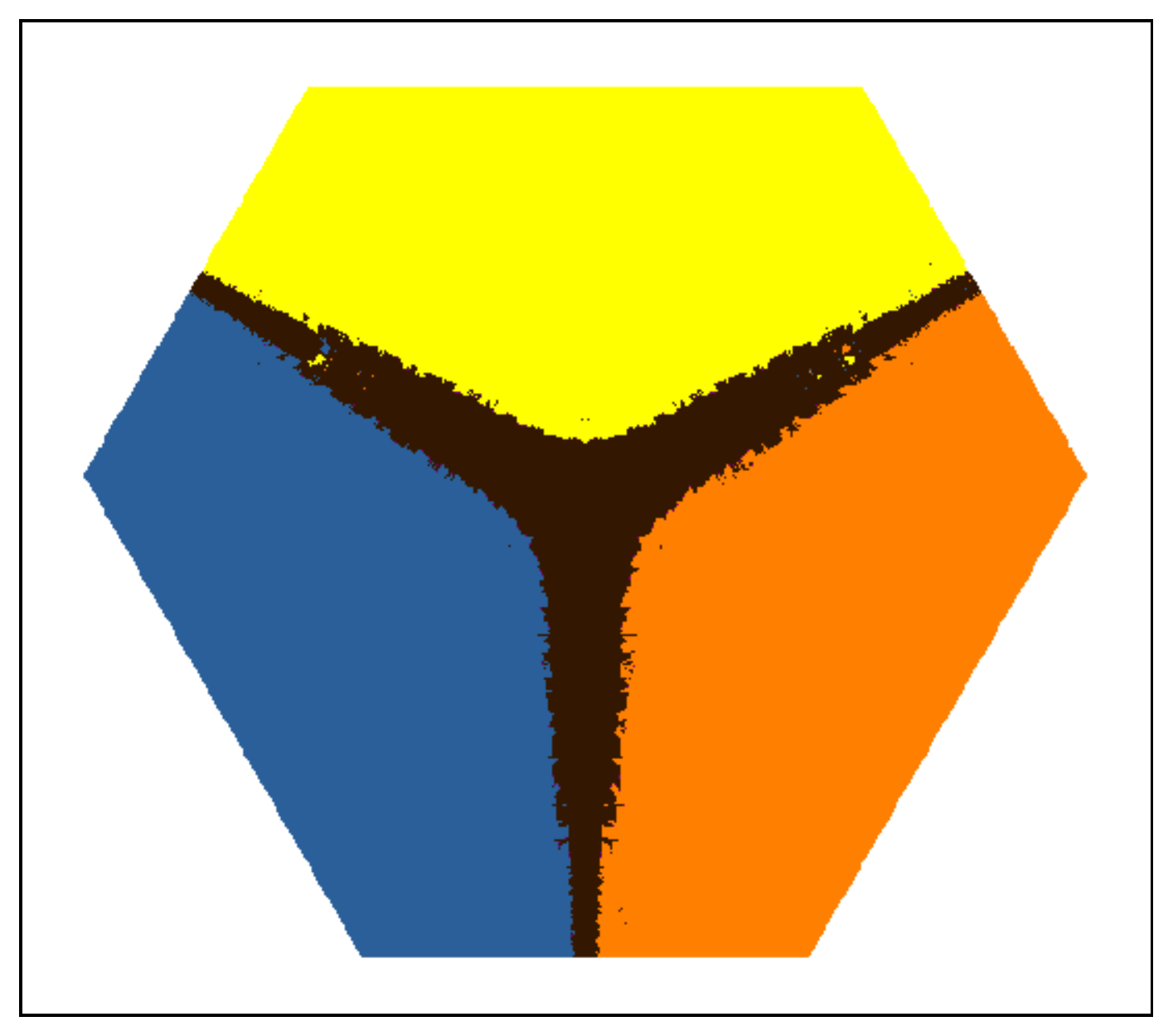} \\
      FIRE, $\Delta=0.5$ \\ 
      \includegraphics[width=0.4\textwidth]{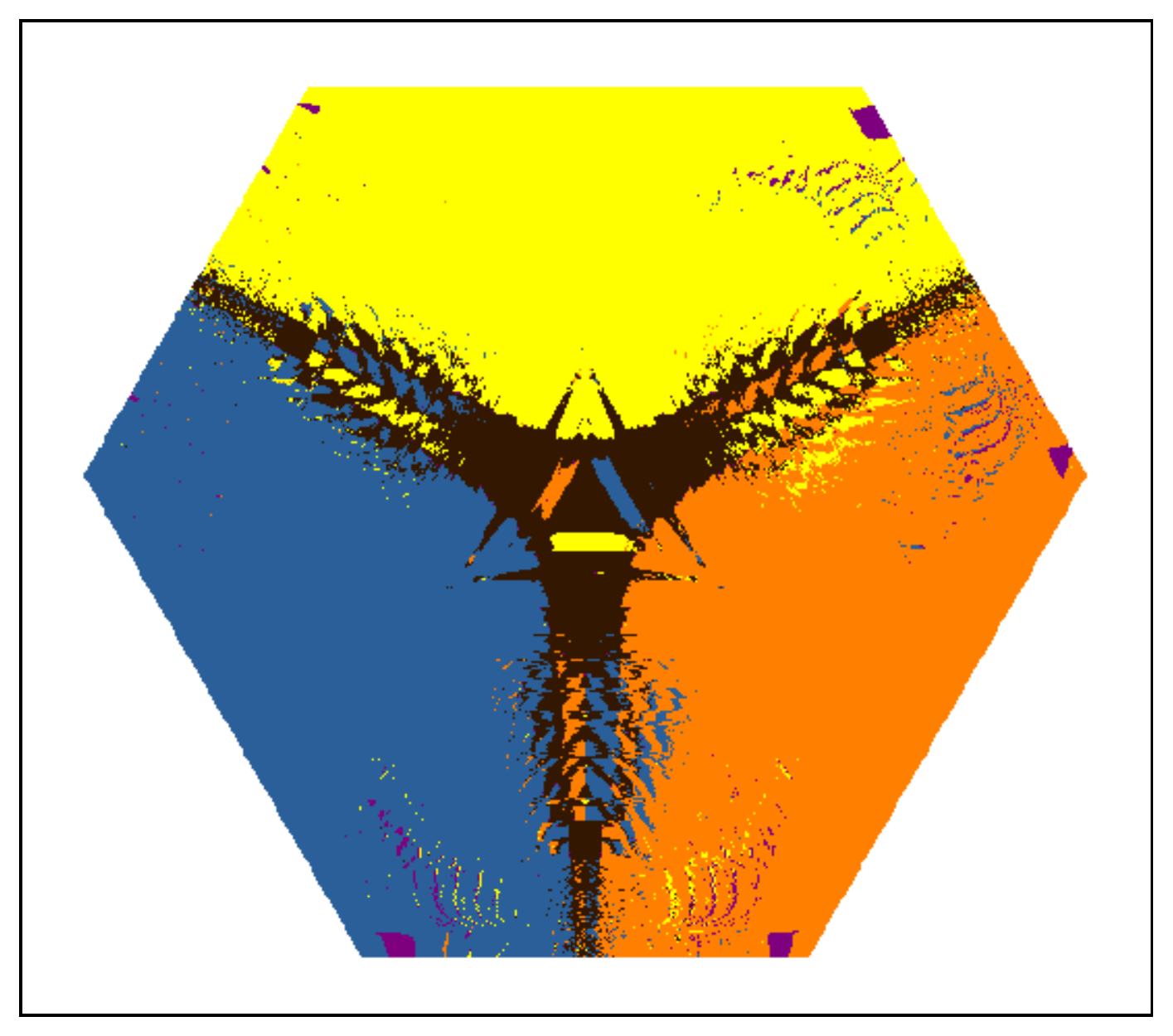}
      \includegraphics[width=0.4\textwidth]{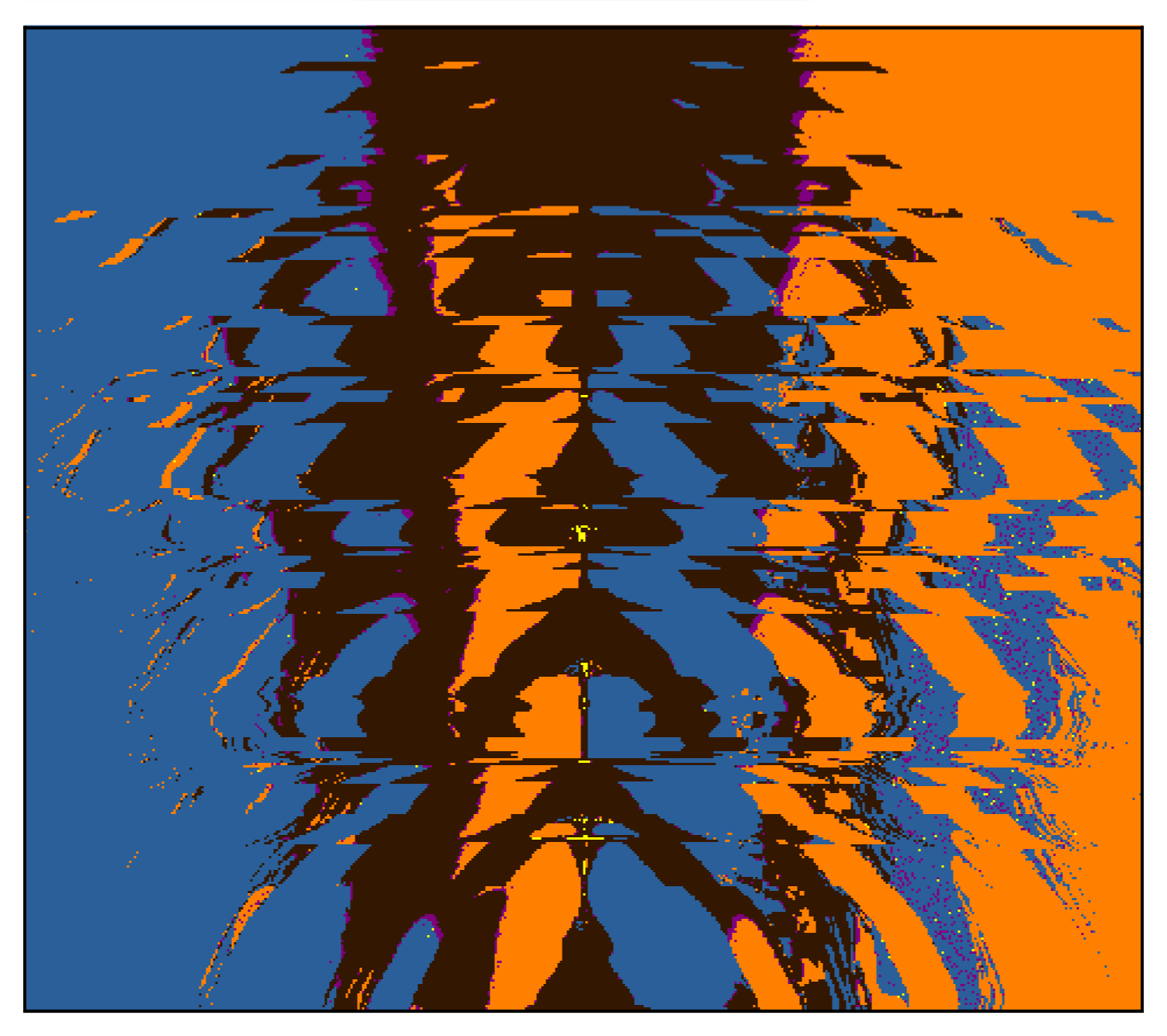} 
      \caption{Results for the FIRE algorithm for several values of the maximum
        step size $\Delta$.  The bottom right panel is a magnification of a
        portion of the bottom left image.  \seesd } 
      \label{fig:fire}
\end{figure*}

Several values of the maximum step size for FIRE were tested. For a
small value of the step size the boundaries of the basins of
attraction are well defined and similar to the results for
steepest-descent.  The method only ends up in the wrong basin when
starting from points that lie very close to the boundaries between two
basins (\ref{fig:fire}, top left).  Using a larger value of the step
size leads to many artifacts and failed quenches, which are evident in
the bottom half of \ref{fig:fire}.

\begin{figure*}
  Conjugate gradient, with line search (SciPy) \\
    \includegraphics[width=0.4\textwidth]{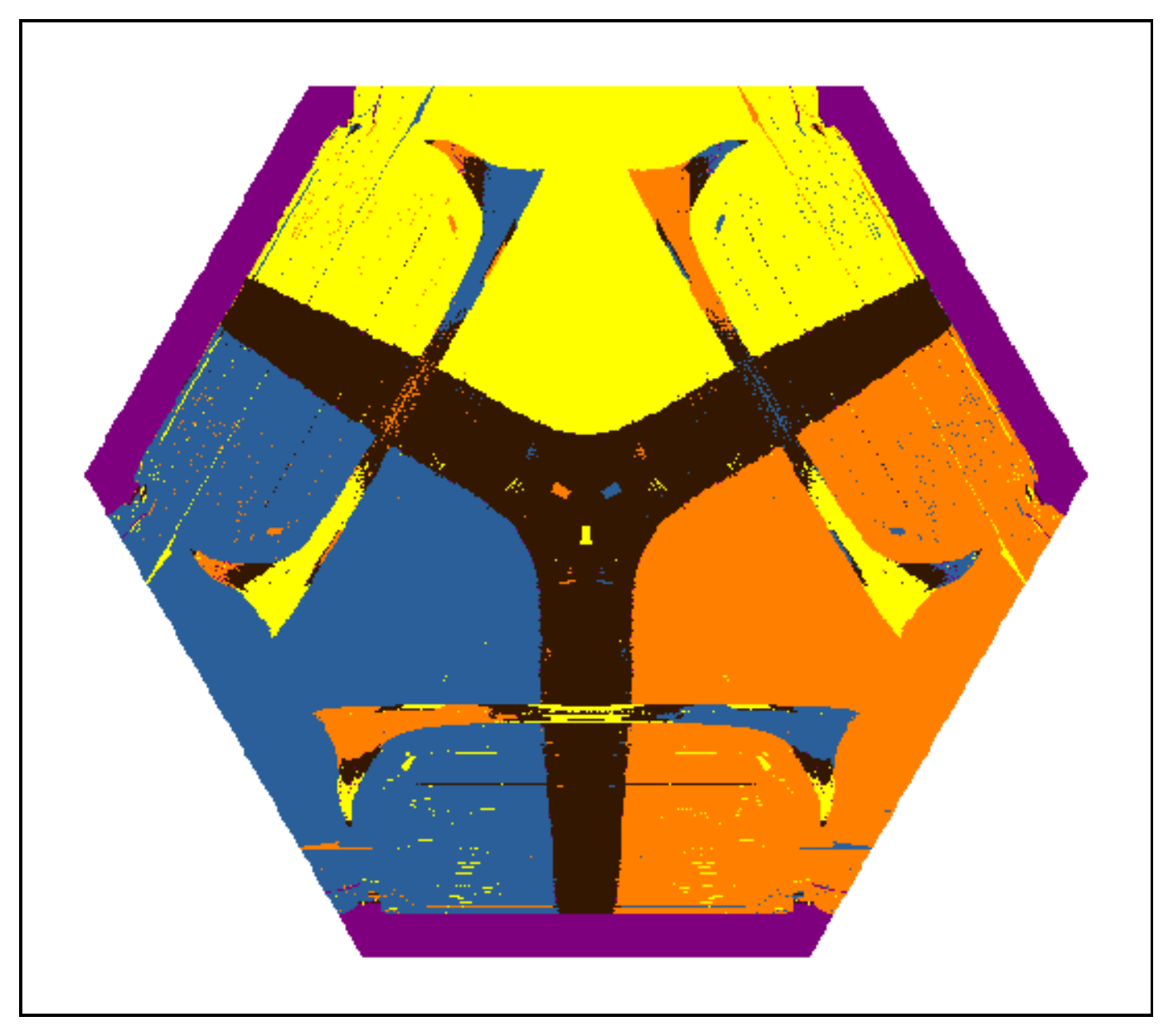}
    \includegraphics[width=0.4\textwidth]{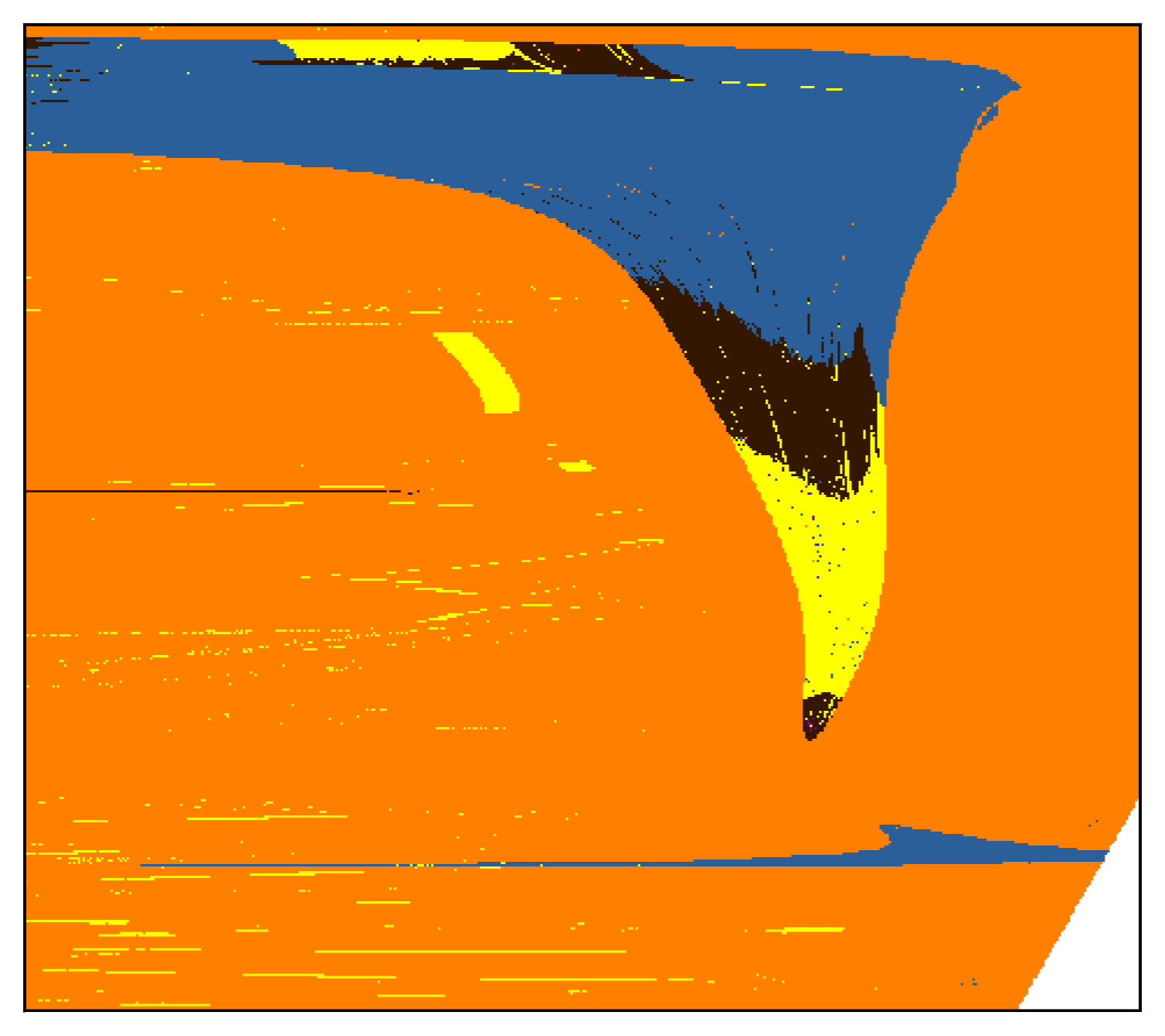}
    \caption{Results for the conjugate gradient algorithm with line search
      (SciPy implementation) are shown in the left panel. The right panel shows
      a magnification of the left panel.  
      \seesd}
    \label{fig:cg}
\end{figure*}

Some other popular algorithms were also tested, namely, conjugate
gradient (\ref{fig:cg}) and BFGS (\ref{fig:bfgs}). The SciPy
implementation of these methods uses the same line search routine to
determine the step size at each iteration, and both of them produce
similar ill-defined basin boundaries.  In both cases, the boundary
artifacts are caused by the line search returning a step size that is
large enough to move into a different basin.  Furthermore, the failed
quenches at the edges are due to step sizes sufficiently large that
particles end up so far apart that the gradient is small enough to
satisfy the termination condition.  The line search algorithm is not
entirely responsible for the imprecise basin boundaries. The initial
guess for the step size passed to the line search by the conjugate
gradient and BFGS algorithms is often large enough to step to the next
basin by itself.  To check this effect, we have also tested this line
search routine with the L-BFGS algorithm.  The results (not shown for
brevity) produce quite reasonable basin boundaries with most of the
above artifacts absent.  An interesting question is why the L-BFGS
algorithm produces an accurate guess for the step size while BFGS
tends to overestimate the step size.  The answer, most likely, is that
the initial Hessian in L-BFGS is scaled,\cite{Nocedal1989} while in
BFGS it is fixed to unity.

We wanted to test the effect of maximum step size on the conjugate
gradient and BFGS algorithms; however the SciPy routines do not accept
parameters for adjusting the maximum step size.  We were able to
introduce this adjustment by modifying the source code of the line
search routine used in each case.  With these modifications and a
maximum step size of $0.1$, the conjugate gradient routine produced
reasonably accurate basin boundaries.  The penalty for this improved
precision was roughly $50\%$ more function evaluations.  We were not
able to obtain significant improvements for the BFGS routine.

The basins of attraction determined by some of the methods mentioned
above (in particular FIRE with a large step size, conjugate gradient
and BFGS) display complex structures, as shown in \ref{fig:fire}
(bottom right) and \ref{fig:cg} (right). Here we can see that the
basin boundaries still have structure, even as the length scale is
reduced.  As noted in previous work, the structure may be fractal,
\cite{Grebogi1987,Wales1992,Wales1993} although this possibility was
not investigated in detail.

\begin{table}
  \caption{\label{tab:difpix} Table showing the quantitative
    analysis of the basins of attraction defined by different
    minimisers for the three-body system. The third
    column, Err, is the percentage of starting configurations that minimised to  
    a different minimum compared to the steepest-descent method. $\Delta$ is
    the maximum step size and $M$ is the length of the history used by
    L-BFGS, as defined in the Methods section.}
    \begin{tabular}{llr}
    \hline \hline
      {\bf Algorithm} & {\bf Parameters} & {\bf Err (\%)} \\
      \hline \hline
      
      {\bf L-BFGS} & $M = 1$, $\phantom{0}\Delta = 0.05$ & 6.41 \\
      (without LS)& $M = 1$, $\phantom{0}\Delta = 0.1$ & 6.46 \\
      & $M = 1$, $\phantom{0}\Delta = 0.2$ & 7.16 \\
      
      & $M = 4$, $\phantom{0}\Delta = 0.05$ & 14.75 \\
      & $M = 4$, $\phantom{0}\Delta = 0.1$ & 10.84 \\
      & $M = 4$, $\phantom{0}\Delta = 0.2$ & 10.13 \\
      
      & $M = 10$, $\Delta = 0.05$ & 16.78 \\
      & $M = 10$, $\Delta = 0.1$ & 12.72 \\
      & $M = 10$, $\Delta = 0.2$ & 14.81 \\
      
      \hline
      {\bf L-BFGS} & & \\
      (SciPy, with LS) & $M = 10$ &  12.79 \\
      \hline
      {\bf FIRE} &$\Delta = 0.1$&  3.68 \\
      & $\Delta = 0.25$ &  4.53 \\
      & $\Delta = 0.5$ &  11.23 \\
      \hline
      {\bf BFGS} && \\
      (SciPy, with LS)& &  23.45 \\
      \hline
      {\bf CG} && \\
      (SciPy, with LS)&&  22.93 \\
      \hline
      {\bf Steepest-Descent} & $\Delta = 0.001$ & 0.00 \\
      \hline \hline
      \end{tabular}
\end{table}

We have quantified the difference between the outcomes of the basin
mapping for different minimisers by counting the number of starting
structures for which we find a basin different from the one obtained
using steepest-descent, as shown in \ref{tab:difpix}. This difference
corresponds to the number of different structures (from a total of 170,607
valid starting structures) when comparing the minimum produced by the
corresponding algorithm with the minimum produced by steepest-descent
(\ref{fig:sd}).

\begin{table}
  \caption{\label{tab:err} Benchmarks for the algorithms tested in this paper
    in terms of the average number of function calls
    $\langle$\text{FCs}$\rangle$ needed to minimise a 38-particle Lennard-Jones
    system from a random configuration.  The stopping condition is that the
    maximum force on any atom is less than 0.01 in reduced units.}
    \begin{tabular}{llr}
    \hline \hline
      {\bf Algorithm} & {\bf Parameters} & $\langle$\text{\bf FCs}$\rangle$\\
      \hline \hline
      
      {\bf L-BFGS}& $M = 1$, $\phantom{0}\Delta = 0.1$ & 273 \\ 
      (without LS)& $M = 4$, $\phantom{0}\Delta = 0.05$& 369 \\
      & $M = 4$, $\phantom{0}\Delta = 0.1$ & 241 \\
      & $M = 4$, $\phantom{0}\Delta = 0.2$ & 225 \\
      & $M = 10$, $\Delta = 0.1$ & 228 \\
      & $M = 20$, $\Delta = 0.1$ & 216 \\
      \hline
      {\bf L-BFGS} & & \\
      (SciPy, with LS) & $M = 10$ & 215 \\
      \hline
      {\bf FIRE} &$\Delta = 0.1$& 822 \\
      & $\Delta = 0.5$ & 3,185 \\
      \hline
      {\bf BFGS} && \\
      (SciPy, with LS)&& 1,233 \\
      \hline
      {\bf CG} && \\
      (SciPy, with LS)&& 837 \\
      \hline
      {\bf Steepest-Descent} & $\Delta = 0.001$ & 31,672\\
      \hline \hline
      \end{tabular}
\end{table}

\ref{tab:err} reports the performance of the algorithms tested here in
terms of the average number of times that the energy and the force
were evaluated $\langle$\text{FCs}$\rangle$. The average is taken over
a sample of 1,000 random initial states for a 38-particle
Lennard-Jones cluster. The number of evaluations ultimately determines
the time it takes to find a minimum, as this is generally the most
time consuming part of any minimisation algorithm. We can see that
L-BFGS is the fastest and FIRE is about three to four times slower,
while steepest-descent is orders of magnitude slower, as expected.

\section{Conclusions}
\label{sec:con}

In this paper, we have mapped the basins of attraction of a simple
system onto a plane to compare a number of minimisation algorithms. We
are able to compare the different approaches both visually and
quantitatively, building upon previous work, where the focus was
mainly on transition state searches.\cite{Wales1992,Wales1993} Some of
the more complex algorithms (CG, BFGS, L-BFGS and FIRE) depending on
the choice of parameters produce basins that consist of disconnected
parts. Such basins deviate from the ``correct'' basin of attraction
defined by steepest-descent pathways, especially at the basin
boundaries, where complex interpenetrating patterns can appear. These
patterns generally do not disappear as the length scale is reduced, as
can be seen in \ref{fig:fire} and \ref{fig:cg}, making the basins
ill-defined.  In particular, we have found that overestimates for the
step size are primarily responsible for the complex basin boundaries.
Imposing a maximum step size can mitigate this problem for some
algorithms, at the cost of slightly higher computational effort and an
additional, system dependent, parameter.  An appropriate value for the
maximum step size can be chosen based on length scales in the system:
for atomic systems, a good choice is about one tenth of the
inter-atomic pair equilibrium distance.

In conclusion, if assignment of a starting configuration to the basin
of attraction defined by steepest-descent is important, then FIRE may
be the most convenient algorithm, due to its speed and precision,
provided that an acceptable maximum step size is chosen. If finding a
minimum quickly is more important, then L-BFGS is clearly the best
choice.

\begin{acknowledgement}
  The authors thank Dr Victor R\"uhle for many useful discussions and
  acknowledge funding from EPSRC Programme Grants EP/I001352/1 and
  EP/I00844/1, ERC Advanced Grants RG59508 and 227758, Marie Curie
  Grant 275544, Wolfson Merit Award 502011.K701/JE and Becas Chile
  CONICYT.
\end{acknowledgement}

\bibliography{compareminmeth}

\providecommand*{\mcitethebibliography}{\thebibliography}
\csname @ifundefined\endcsname{endmcitethebibliography}
{\let\endmcitethebibliography\endthebibliography}{}
\begin{mcitethebibliography}{40}
\providecommand*{\natexlab}[1]{#1}
\providecommand*{\mciteSetBstSublistMode}[1]{}
\providecommand*{\mciteSetBstMaxWidthForm}[2]{}
\providecommand*{\mciteBstWouldAddEndPuncttrue}
  {\def\EndOfBibitem{\unskip.}}
\providecommand*{\mciteBstWouldAddEndPunctfalse}
  {\let\EndOfBibitem\relax}
\providecommand*{\mciteSetBstMidEndSepPunct}[3]{}
\providecommand*{\mciteSetBstSublistLabelBeginEnd}[3]{}
\providecommand*{\EndOfBibitem}{}
\mciteSetBstSublistMode{f}
\mciteSetBstMaxWidthForm{subitem}{(\alph{mcitesubitemcount})}
\mciteSetBstSublistLabelBeginEnd{\mcitemaxwidthsubitemform\space}
{\relax}{\relax}

\bibitem[Wales(2003)]{Wales2003}
Wales,~D.~J. \emph{Energy Landscapes: With Applications to Clusters,
  Biomolecules and Glasses};
\newblock Cambridge University Press: Cambridge, U.K., 2003\relax
\mciteBstWouldAddEndPuncttrue
\mciteSetBstMidEndSepPunct{\mcitedefaultmidpunct}
{\mcitedefaultendpunct}{\mcitedefaultseppunct}\relax
\EndOfBibitem
\bibitem[Murrell and Laidler(1968)]{murrelll68}
Murrell,~J.~N.; Laidler,~K.~J. \emph{Trans. Faraday. Soc.} \textbf{1968},
  \emph{64}, 371--377\relax
\mciteBstWouldAddEndPuncttrue
\mciteSetBstMidEndSepPunct{\mcitedefaultmidpunct}
{\mcitedefaultendpunct}{\mcitedefaultseppunct}\relax
\EndOfBibitem
\bibitem[Kirkpatrick et~al.(1989)Kirkpatrick, Thirumalai, and
  Wolynes]{kirkpatricktw89}
Kirkpatrick,~T.~R.; Thirumalai,~D.; Wolynes,~P.~G. \emph{Phys. Rev. A}
  \textbf{1989}, \emph{40}, 1045\relax
\mciteBstWouldAddEndPuncttrue
\mciteSetBstMidEndSepPunct{\mcitedefaultmidpunct}
{\mcitedefaultendpunct}{\mcitedefaultseppunct}\relax
\EndOfBibitem
\bibitem[Doliwa and Heuer(2003)]{DoliwaH03}
Doliwa,~B.; Heuer,~A. \emph{Phys. Rev. E} \textbf{2003}, \emph{67},
  031506\relax
\mciteBstWouldAddEndPuncttrue
\mciteSetBstMidEndSepPunct{\mcitedefaultmidpunct}
{\mcitedefaultendpunct}{\mcitedefaultseppunct}\relax
\EndOfBibitem
\bibitem[de~Souza and Wales(2008)]{deSouzaW08}
de~Souza,~V.~K.; Wales,~D.~J. \emph{J. Chem. Phys.} \textbf{2008}, \emph{129},
  164507\relax
\mciteBstWouldAddEndPuncttrue
\mciteSetBstMidEndSepPunct{\mcitedefaultmidpunct}
{\mcitedefaultendpunct}{\mcitedefaultseppunct}\relax
\EndOfBibitem
\bibitem[Wolynes et~al.(1995)Wolynes, Onuchic, and Thirumalai]{Wolynes1995}
Wolynes,~P.; Onuchic,~J.; Thirumalai,~D. \emph{Science} \textbf{1995},
  \emph{267}, 1619--1620\relax
\mciteBstWouldAddEndPuncttrue
\mciteSetBstMidEndSepPunct{\mcitedefaultmidpunct}
{\mcitedefaultendpunct}{\mcitedefaultseppunct}\relax
\EndOfBibitem
\bibitem[Wales and Bogdan(2006)]{WalesB06}
Wales,~D.~J.; Bogdan,~T.~V. \emph{J. Phys. Chem. B} \textbf{2006}, \emph{110},
  20765--20776\relax
\mciteBstWouldAddEndPuncttrue
\mciteSetBstMidEndSepPunct{\mcitedefaultmidpunct}
{\mcitedefaultendpunct}{\mcitedefaultseppunct}\relax
\EndOfBibitem
\bibitem[Mehta et~al.(2013)Mehta, Hauenstein, and Wales]{MehtaHW13}
Mehta,~D.; Hauenstein,~J.~D.; Wales,~D.~J. \emph{J. Chem. Phys. Commun,}
  \textbf{2013}, \emph{000}, 000000\relax
\mciteBstWouldAddEndPuncttrue
\mciteSetBstMidEndSepPunct{\mcitedefaultmidpunct}
{\mcitedefaultendpunct}{\mcitedefaultseppunct}\relax
\EndOfBibitem
\bibitem[Wales(1993)]{wales93f}
Wales,~D.~J. \emph{Mol. Phys.} \textbf{1993}, \emph{78}, 151--171\relax
\mciteBstWouldAddEndPuncttrue
\mciteSetBstMidEndSepPunct{\mcitedefaultmidpunct}
{\mcitedefaultendpunct}{\mcitedefaultseppunct}\relax
\EndOfBibitem
\bibitem[Doye and Wales(1995)]{doyew95c}
Doye,~J. P.~K.; Wales,~D.~J. \emph{J. Chem. Phys.} \textbf{1995}, \emph{102},
  9673--9688\relax
\mciteBstWouldAddEndPuncttrue
\mciteSetBstMidEndSepPunct{\mcitedefaultmidpunct}
{\mcitedefaultendpunct}{\mcitedefaultseppunct}\relax
\EndOfBibitem
\bibitem[Strodel and Wales(2008)]{StrodelW08b}
Strodel,~B.; Wales,~D.~J. \emph{Chem. Phys. Lett.} \textbf{2008}, \emph{466},
  105--115\relax
\mciteBstWouldAddEndPuncttrue
\mciteSetBstMidEndSepPunct{\mcitedefaultmidpunct}
{\mcitedefaultendpunct}{\mcitedefaultseppunct}\relax
\EndOfBibitem
\bibitem[Mezey(1981)]{mezey81b}
Mezey,~P.~G. \emph{Theo. Chim. Acta} \textbf{1981}, \emph{58}, 309\relax
\mciteBstWouldAddEndPuncttrue
\mciteSetBstMidEndSepPunct{\mcitedefaultmidpunct}
{\mcitedefaultendpunct}{\mcitedefaultseppunct}\relax
\EndOfBibitem
\bibitem[Pechukas(1976)]{pechukas76}
Pechukas,~P. \emph{J. Chem. Phys.} \textbf{1976}, \emph{64}, 1516\relax
\mciteBstWouldAddEndPuncttrue
\mciteSetBstMidEndSepPunct{\mcitedefaultmidpunct}
{\mcitedefaultendpunct}{\mcitedefaultseppunct}\relax
\EndOfBibitem
\bibitem[Wales(1992)]{Wales1992}
Wales,~D.~J. \emph{J. Chem. Soc., Faraday Trans.} \textbf{1992}, \emph{88},
  653\relax
\mciteBstWouldAddEndPuncttrue
\mciteSetBstMidEndSepPunct{\mcitedefaultmidpunct}
{\mcitedefaultendpunct}{\mcitedefaultseppunct}\relax
\EndOfBibitem
\bibitem[Wales(1993)]{Wales1993}
Wales,~D.~J. \emph{J. Chem. Soc., Faraday Trans.} \textbf{1993}, \emph{89},
  1305\relax
\mciteBstWouldAddEndPuncttrue
\mciteSetBstMidEndSepPunct{\mcitedefaultmidpunct}
{\mcitedefaultendpunct}{\mcitedefaultseppunct}\relax
\EndOfBibitem
\bibitem[Wales and Doye(1997)]{walesd97a}
Wales,~D.~J.; Doye,~J. P.~K. \emph{J. Phys. Chem. A} \textbf{1997}, \emph{101},
  5111--5116\relax
\mciteBstWouldAddEndPuncttrue
\mciteSetBstMidEndSepPunct{\mcitedefaultmidpunct}
{\mcitedefaultendpunct}{\mcitedefaultseppunct}\relax
\EndOfBibitem
\bibitem[Li and Scheraga(1987)]{lis87}
Li,~Z.; Scheraga,~H.~A. \emph{Proc. Natl. Acad. Sci. U.S.A.} \textbf{1987},
  \emph{84}, 6611\relax
\mciteBstWouldAddEndPuncttrue
\mciteSetBstMidEndSepPunct{\mcitedefaultmidpunct}
{\mcitedefaultendpunct}{\mcitedefaultseppunct}\relax
\EndOfBibitem
\bibitem[Xu et~al.(2011)Xu, Frenkel, and Liu]{xu13}
Xu,~N.; Frenkel,~D.; Liu,~A.~J. \emph{Phys. Rev. Lett.} \textbf{2011},
  \emph{106}, 245502\relax
\mciteBstWouldAddEndPuncttrue
\mciteSetBstMidEndSepPunct{\mcitedefaultmidpunct}
{\mcitedefaultendpunct}{\mcitedefaultseppunct}\relax
\EndOfBibitem
\bibitem[Nocedal and Wright(1999)]{Nocedal1999}
Nocedal,~J.; Wright,~S. \emph{Numerical Optimization};
\newblock Springer Science+Business Media: New York, NY, U.S.A., 1999\relax
\mciteBstWouldAddEndPuncttrue
\mciteSetBstMidEndSepPunct{\mcitedefaultmidpunct}
{\mcitedefaultendpunct}{\mcitedefaultseppunct}\relax
\EndOfBibitem
\bibitem[Wales()]{gmin}
Wales,~D.~J. \emph{GMIN: A Program for Finding Global Minima and Calculating
  Thermodynamic Properties from Basin-sampling.}
  \url{http://www-wales.ch.cam.ac.uk/GMIN/}\relax
\mciteBstWouldAddEndPuncttrue
\mciteSetBstMidEndSepPunct{\mcitedefaultmidpunct}
{\mcitedefaultendpunct}{\mcitedefaultseppunct}\relax
\EndOfBibitem
\bibitem[Wales()]{optim}
Wales,~D.~J. \emph{OPTIM: A Program for Optimising Geometries and Calculating
  Pathways}\relax
\mciteBstWouldAddEndPuncttrue
\mciteSetBstMidEndSepPunct{\mcitedefaultmidpunct}
{\mcitedefaultendpunct}{\mcitedefaultseppunct}\relax
\EndOfBibitem
\bibitem[Broyden(1970)]{Broyden1970}
Broyden,~C.~G. \emph{IMA J. Appl. Math.} \textbf{1970}, \emph{6}, 76--90\relax
\mciteBstWouldAddEndPuncttrue
\mciteSetBstMidEndSepPunct{\mcitedefaultmidpunct}
{\mcitedefaultendpunct}{\mcitedefaultseppunct}\relax
\EndOfBibitem
\bibitem[Fletcher(1970)]{Fletcher1970}
Fletcher,~R. \emph{Comput. J.} \textbf{1970}, \emph{13}, 317--322\relax
\mciteBstWouldAddEndPuncttrue
\mciteSetBstMidEndSepPunct{\mcitedefaultmidpunct}
{\mcitedefaultendpunct}{\mcitedefaultseppunct}\relax
\EndOfBibitem
\bibitem[Goldfarb(1970)]{Goldfarb1970}
Goldfarb,~D. \emph{Math. Comp.} \textbf{1970}, \emph{24}, 23--26\relax
\mciteBstWouldAddEndPuncttrue
\mciteSetBstMidEndSepPunct{\mcitedefaultmidpunct}
{\mcitedefaultendpunct}{\mcitedefaultseppunct}\relax
\EndOfBibitem
\bibitem[Shanno(1970)]{Shanno1970}
Shanno,~D.~F. \emph{Math. Comp.} \textbf{1970}, \emph{24}, 647--656\relax
\mciteBstWouldAddEndPuncttrue
\mciteSetBstMidEndSepPunct{\mcitedefaultmidpunct}
{\mcitedefaultendpunct}{\mcitedefaultseppunct}\relax
\EndOfBibitem
\bibitem[Jones et~al.(2001--)Jones, Oliphant, and Peterson]{scipy}
Jones,~E.; Oliphant,~T.; Peterson,~P. \emph{{SciPy}: Open source Scientific
  Tools for {Python}}, 2001--. \url{http://www.scipy.org/}\relax
\mciteBstWouldAddEndPuncttrue
\mciteSetBstMidEndSepPunct{\mcitedefaultmidpunct}
{\mcitedefaultendpunct}{\mcitedefaultseppunct}\relax
\EndOfBibitem
\bibitem[More and Thuente(1994)]{more.1994}
More,~J.~J.; Thuente,~D.~J. \emph{Acm T. Math. Software} \textbf{1994},
  \emph{20}, 286--307\relax
\mciteBstWouldAddEndPuncttrue
\mciteSetBstMidEndSepPunct{\mcitedefaultmidpunct}
{\mcitedefaultendpunct}{\mcitedefaultseppunct}\relax
\EndOfBibitem
\bibitem[Liu and Nocedal(1989)]{Nocedal1989}
Liu,~D.~C.; Nocedal,~J. \emph{Math. Program.} \textbf{1989}, \emph{45},
  503--528\relax
\mciteBstWouldAddEndPuncttrue
\mciteSetBstMidEndSepPunct{\mcitedefaultmidpunct}
{\mcitedefaultendpunct}{\mcitedefaultseppunct}\relax
\EndOfBibitem
\bibitem[Zhu et~al.(1997)Zhu, Byrd, Lu, and Nocedal]{Zhu1997}
Zhu,~C.; Byrd,~R.~H.; Lu,~P.; Nocedal,~J. \emph{{ACM} Trans. Math. Software}
  \textbf{1997}, \emph{23}, 550--560\relax
\mciteBstWouldAddEndPuncttrue
\mciteSetBstMidEndSepPunct{\mcitedefaultmidpunct}
{\mcitedefaultendpunct}{\mcitedefaultseppunct}\relax
\EndOfBibitem
\bibitem[Byrd et~al.(1995)Byrd, Lu, Nocedal, and Zhu]{Byrd1995}
Byrd,~R.; Lu,~P.; Nocedal,~J.; Zhu,~C. \emph{SIAM J. Sci. Comput.}
  \textbf{1995}, \emph{16}, 1190--1208\relax
\mciteBstWouldAddEndPuncttrue
\mciteSetBstMidEndSepPunct{\mcitedefaultmidpunct}
{\mcitedefaultendpunct}{\mcitedefaultseppunct}\relax
\EndOfBibitem
\bibitem[Morales and Nocedal(2011)]{Morales2011}
Morales,~J.~L.; Nocedal,~J. \emph{{ACM} Trans. Math. Software} \textbf{2011},
  \emph{38}, 7:1--7:4\relax
\mciteBstWouldAddEndPuncttrue
\mciteSetBstMidEndSepPunct{\mcitedefaultmidpunct}
{\mcitedefaultendpunct}{\mcitedefaultseppunct}\relax
\EndOfBibitem
\bibitem[Bitzek et~al.(2006)Bitzek, Koskinen, G\"ahler, Moseler, and
  Gumbsch]{Bitzek2006}
Bitzek,~E.; Koskinen,~P.; G\"ahler,~F.; Moseler,~M.; Gumbsch,~P. \emph{Phys.
  Rev. Lett.} \textbf{2006}, \emph{97}, 170201\relax
\mciteBstWouldAddEndPuncttrue
\mciteSetBstMidEndSepPunct{\mcitedefaultmidpunct}
{\mcitedefaultendpunct}{\mcitedefaultseppunct}\relax
\EndOfBibitem
\bibitem[Hestenes and Stiefel(1952)]{Stiefel1952}
Hestenes,~M.~R.; Stiefel,~E. \emph{J. Res. Nat. Bur. Stand.} \textbf{1952},
  \emph{49}, 409--436\relax
\mciteBstWouldAddEndPuncttrue
\mciteSetBstMidEndSepPunct{\mcitedefaultmidpunct}
{\mcitedefaultendpunct}{\mcitedefaultseppunct}\relax
\EndOfBibitem
\bibitem[Lennard-Jones(1924)]{LJ1924}
Lennard-Jones,~J.~E. \emph{Proc. R. Soc. A} \textbf{1924}, \emph{106},
  463--477\relax
\mciteBstWouldAddEndPuncttrue
\mciteSetBstMidEndSepPunct{\mcitedefaultmidpunct}
{\mcitedefaultendpunct}{\mcitedefaultseppunct}\relax
\EndOfBibitem
\bibitem[Axilrod and Teller(1943)]{Axilrod1943}
Axilrod,~B.~M.; Teller,~E. \emph{J. Chem. Phys.} \textbf{1943}, \emph{11},
  299--300\relax
\mciteBstWouldAddEndPuncttrue
\mciteSetBstMidEndSepPunct{\mcitedefaultmidpunct}
{\mcitedefaultendpunct}{\mcitedefaultseppunct}\relax
\EndOfBibitem
\bibitem[Wales(1990)]{wales90a}
Wales,~D.~J. \emph{J. Chem. Soc., Faraday Trans.} \textbf{1990}, \emph{86},
  3505--3517\relax
\mciteBstWouldAddEndPuncttrue
\mciteSetBstMidEndSepPunct{\mcitedefaultmidpunct}
{\mcitedefaultendpunct}{\mcitedefaultseppunct}\relax
\EndOfBibitem
\bibitem[Doye and Wales(1992)]{doyew92}
Doye,~J. P.~K.; Wales,~D.~J. \emph{J. Chem. Soc., Faraday Trans.}
  \textbf{1992}, \emph{88}, 3295--3304\relax
\mciteBstWouldAddEndPuncttrue
\mciteSetBstMidEndSepPunct{\mcitedefaultmidpunct}
{\mcitedefaultendpunct}{\mcitedefaultseppunct}\relax
\EndOfBibitem
\bibitem[Uppenbrink and Wales(1992)]{uppenbrinkw92b}
Uppenbrink,~J.; Wales,~D.~J. \emph{Chem. Phys. Lett.} \textbf{1992},
  \emph{190}, 447--452\relax
\mciteBstWouldAddEndPuncttrue
\mciteSetBstMidEndSepPunct{\mcitedefaultmidpunct}
{\mcitedefaultendpunct}{\mcitedefaultseppunct}\relax
\EndOfBibitem
\bibitem[Grebogi et~al.(1987)Grebogi, Ott, and Yorke]{Grebogi1987}
Grebogi,~C.; Ott,~E.; Yorke,~J.~A. \emph{Science} \textbf{1987}, \emph{238},
  632--638\relax
\mciteBstWouldAddEndPuncttrue
\mciteSetBstMidEndSepPunct{\mcitedefaultmidpunct}
{\mcitedefaultendpunct}{\mcitedefaultseppunct}\relax
\EndOfBibitem
\end{mcitethebibliography}


\end{document}